\documentclass[preprintnumbers, prd, showpacs, floatfix, preprintnumbers, letterpaper, nofootinbib, amsmath, amssymb, superscriptaddress,preprint]
{revtex4-1}
\usepackage{graphicx}
\usepackage{epsfig}
\usepackage{bm}
\usepackage{amsfonts}

\usepackage{color}

\newbox\pippobox

\def\be{\begin{equation}}
\def\ee{\end{equation}}
\def\ba{\begin{eqnarray} }
\def\ea{\end{eqnarray}}

\newcommand {\lla} {\ {\raise-.5ex\hbox{$\buildrel<\over\sim$}}\ }

\usepackage[T1]{fontenc}
\usepackage[latin1]{inputenc}
\usepackage{graphicx}
\usepackage[english]{babel}
\usepackage{amsmath}
\usepackage{amssymb}
\usepackage{amsfonts}

\def\be{\begin{equation}}
\def\ee{\end{equation}}
\def\bea{\begin{eqnarray}}
\def\eea{\end{eqnarray}}

\begin{document}

\title{A class of black holes in dRGT massive gravity and their thermodynamical properties}

\author{Sushant G. Ghosh \footnote{Email: sghosh2@jmi.ac.in, sgghosh@gmail.com }}
\affiliation{Centre of Theoretical Physics, Jamia Millia Islamia, New Delhi 110025, India}
\affiliation{Astrophysics and Cosmology Research Unit, School of Mathematical Sciences, University of Kwazulu-Natal, Private Bag 54001, Durban 4000, South Africa}

\author{Lunchakorn Tannukij \footnote{Email: l\_tannukij@hotmail.com}}
\affiliation{ Department of Physics, Faculty of Science,  Mahidol University, Bangkok 10400, Thailand}

\author{Pitayuth Wongjun \footnote{Email: pitbaa@gmail.com}}
\affiliation{The institute for fundamental study, Naresuan University, Phitsanulok 65000, Thailand}
\affiliation{Thailand Center of Excellence in Physics, Ministry of Education,
Bangkok 10400, Thailand}

\begin{abstract}
We present an exact spherical black hole solution in de Rham, Gabadadze and Tolley  (dRGT) massive gravity for a generic choice of the parameters in the theory, and also discuss the thermodynamical and phase structure of the black hole in both the grand canonical and canonical ensembles (for the charged case). It turns out that the dRGT black hole solution includes other known solutions to the Einstein field equations, such as the monopole-de Sitter-Schwarzschild solution with the coefficients of the third and fourth terms in the
potential and the graviton mass in massive gravity naturally generates the cosmological constant and the global monopole term.
Furthermore, we compute the mass, temperature and entropy of the dRGT black hole, and also perform thermodynamical stability analysis. It turns out that the presence of the graviton mass completely changes the black hole thermodynamics, and it can provide the Hawking-Page phase transition which also occurs for the charged black holes.    Interestingly, the entropy of a black hole is barely affected and still obeys the standard area law. In particular, our results, in the limit $m_g \rightarrow 0$, reduced exactly  to  \emph{vis-$\grave{a}$-vis}  the general relativity results.
\end{abstract}

\maketitle

\section{Introduction}
The question of whether a mass term for the graviton field can be introduced existed in Einstein's theory of general relativity
since its inception.   Massive gravity came into existence as a straightforward  modification of general relativity  by providing consistent interaction terms which are interpreted as a graviton mass.  Such a theory can describe our Universe, which is currently undergoing accelerating expansion without introducing a bare cosmological constant. Massive gravity modifies gravity by weakening it at the large scale compared with general relativity, which allows the Universe to accelerate, while its predictions at  small scales are the same as those in general relativity.  Furthermore, if solution exists in this theory, it may also elucidate  the dark energy problem.  Hence, in recent years  there were numerous developments in the massive gravity theories \cite{Fierz:1939ix,VanNieuwenhuizen:1973fi,vanDam:1970vg,Zakharov:1970cc,Vainshtein:1972sx,Boulware:1973my}.
The first attempt was done, in 1939, by Fierz and Pauli \cite{Fierz:1939ix}. They added the interaction terms at the linearized level of general relativity but later it was found that their theory suffered from discontinuity in predictions which  was pointed out by van Dam, Veltman, and Zakharov, the so-called van Dam-Veltman-Zakharov (vDVZ) discontinuity \cite{VanNieuwenhuizen:1973fi,vanDam:1970vg,Zakharov:1970cc}. This discontinuity problem invoked further studies on the nonlinear generalization of Fierz-Pauli massive gravity. During the search for such generalizations, Vainshtein found the origin of the vDVZ discontinuity is that the prediction made by the linearized theory cannot be trusted inside some characteristic ``Vainshtein'' radius and he also proposed the mechanism  {for the nonlinear massive gravity} which can be used to recover the predictions made by general relativity  \cite{Vainshtein:1972sx}. At the same time, Boulware and Deser found that such nonlinear generalizations usually generate an equation of motion which has a higher derivative term yielding a ghost instability in the theory, later called Boulware-Deser (BD) ghost \cite{Boulware:1973my}.
However,  these problems, arising in the construction of the massive gravity have been resolved in the last decade by
first introducing St\"uckelberg  fields \cite{hgs}. This permits a class
of potential energies depending on the gravitational metric
and an internal Minkowski metric. Furthermore, to avoid
reappearance of the ghost in massive gravity, the set of
allowed mass terms was confined and furnished perturbatively
by de Rham, Gabadadze and Tolley (dRGT) \cite{deRham:2010ik,deRham:2010kj}.  They summed
these terms and found three possibilities, viz.
quadratic, cubic and quartic combinations of the mass terms.  The dRGT massive gravity is  constructed suitably so that the equations of motion  {have} no higher derivative term, so that the ghost field is
absent.  However, these nonlinear terms lead to complexity  in the calculations and hence, in general, finding
exact solutions in this theory is strenuous. Nevertheless, recently,
several interesting measures have been taken to obtain the spherically symmetric black holes in various massive gravities \cite{Vegh:2013sk,Cai:2014znn,Adams:2014vza,Xu:2015rfa,tmn,bcp,Berezhiani:2011mt,ebaf,msv13,tkn,Babichev:2015xha,Capela:2011mh,ebcd,asjs,ccnp,msv12,bcnp}.  In particular,  a spherically symmetric black hole  solution with a Ricci flat horizon in four dimensional massive gravity
with a negative cosmological constant was obtained by Vegh \cite{Vegh:2013sk}, and was generalized to study the corresponding thermodynamical properties and phase transition structure \cite{Cai:2014znn,Adams:2014vza,Xu:2015rfa}.   The spherically symmetric  solutions for dRGT were also addressed in \cite{tmn,bcp}, the corresponding charged black hole solution was found in \cite{Berezhiani:2011mt} and its bi-gravity
extension, was found in \cite{ebaf}, which includes as particular cases as the previously known spherically symmetric black hole solutions.
 (See \cite{msv13,tkn,Babichev:2015xha}, for reviews on black
holes in massive gravity, see also \cite{Capela:2011mh} for the black hole solution in other classes of massive gravity).

The main purpose of this paper is to present a new class of exact  spherically symmetric black hole solutions including generalization to the charged case in dRGT massive gravity, and also to discuss their thermodynamical properties.  It turns out that the solution discussed in this paper represents a generalization of the Schwarzschild solution that includes most of the known dRGT black hole solutions.  The paper is structured as follows. In the next section, we review dRGT massive gravity. We present the modified equations of motion for dRGT massive gravity and a class of exact black hole solutions in section III. The calculation of the thermodynamical quantities
associated with the dRGT black hole solution and the study of the phase structure of a black hole in the
canonical ensemble approach are the main subject of section IV.  The analyses  in section  IV  are extended for the charged case in the section V, and finally we summarize our results and evoke some perspectives to end the paper in section VI.  We have used units which fix the speed of light and the gravitational constant via $8\pi G = c^4 = 1$.

\section{dRGT massive gravity}\label{model}
We begin by reviewing dRGT massive gravity, which is a well known nonlinear generalization of a massive gravity and is free of the BD ghost by incorporating higher order interaction terms into the Lagrangian.
The dRGT Massive gravity can be represented as Einstein gravity
interacting with a scalar field, and hence its action is the well-known Einstein-Hilbert action plus suitable nonlinear interaction terms as given by \cite{deRham:2010kj}
\begin{eqnarray}\label{action}
 S = \int d^4x \sqrt{-g} \frac{1}{2\kappa^2} \Bigg[ R +m_g^2\,\, {\cal U}(g, \phi^a)\Bigg],
\end{eqnarray}
where $R$ is the Ricci scalar  and ${\cal U}$ is a potential for the graviton which modifies the
gravitational sector with the parameter $m_g$ interpreted as  graviton mass. Moreover, the action is written in a unit such that the Newtonian gravitational constant is unity (thus, $\kappa^2\equiv 8\pi$). The effective potential ${\cal U}$ in four-dimensional spacetime is given by
\begin{eqnarray}\label{potential}
 {\cal U}(g, \phi^a) = {\cal U}_2 + \alpha_3{\cal U}_3 +\alpha_4{\cal U}_4 ,
\end{eqnarray}
in which $\alpha_3$ and $\alpha_4$ are dimensionless free parameters of the theory.
The dependencies of the terms ${\cal U}_2$, ${\cal U}_3$ and ${\cal U}_4$ on the metric $g$ and scalar fields $\phi^a$ are defined as
\begin{eqnarray}
 {\cal U}_2&\equiv&[{\cal K}]^2-[{\cal K}^2] ,\\
 {\cal U}_3&\equiv&[{\cal K}]^3-3[{\cal K}][{\cal K}^2]+2[{\cal K}^3] ,\\
 {\cal U}_4&\equiv&[{\cal K}]^4-6[{\cal K}]^2[{\cal K}^2]+8[{\cal K}][{\cal
K}^3]+3[{\cal K}^2]^2-6[{\cal K}^4],
\end{eqnarray}
where
\begin{eqnarray}
 {\cal K}^\mu_\nu =
\delta^\mu_\nu-\sqrt{g^{\mu\sigma}f_{ab}
\partial_\sigma\phi^a\partial_\nu\phi^b}, \label{K-tensor}
\end{eqnarray}
where $f_{ab}$ is a reference (or fiducial) metric and the rectangular brackets denote the traces,
namely $[{\cal K}]={\cal K}^\mu_\mu$ and $[{\cal K}^n]=({\cal K}^n)^\mu_\mu$. The four scalar fields $\phi^a$ are the St\"uckelberg scalars which are introduced to restore general covariance of the theory. Note that, generally, there are additional mass terms for the theory in higher-dimensional spacetime which are provided explicitly in \ref{A.A}.  One may recognize the interaction terms as symmetric polynomials of $\cal K$;  {for a particular order,} each of the coefficients  {of possible combinations} are chosen so that these terms will not excite higher derivative terms in the equations of motion. Actually, this definition of $\cal K$ is not unique since it is possible to have the same action with a different definition of $\cal K$ --- the alternating action is given in \ref{A.A}.

To proceed further, we choose the unitary gauge $\phi^a=x^\mu\delta^a_\mu$ \cite{Vegh:2013sk}. In this gauge, the tensor $g_{\mu\nu}$ is the observable metric describing the five degrees of freedom of the massive graviton. Note that since the St\"uckelberg scalars transform according to the coordinate transformation, once the scalars are fixed, for example,  {due to choosing the} unitary gauge, applying a coordinate transformation will break the gauge  {condition} and then introduce additional changes in the St\"uckelberg scalars. Also, we redefine the two
parameters $\alpha_3$ and $\alpha_4$ of the graviton potential in Eq. \eqref{potential} by introducing two new parameters $\alpha$ and
$\beta$, as follows
\begin{eqnarray}\label{alphabeta}
 \alpha_3 = \frac{\alpha-1}{3}~,~~\alpha_4 =
\frac{\beta}{4}+\frac{1-\alpha}{12}.
\end{eqnarray}

By varying the action with respect to metric $g_{\mu\nu}$, we obtain the modified Einstein field equations as
\begin{eqnarray}\label{EoM}
 G_{\mu\nu} +m_g^2 X_{\mu\nu} = 0, \label{modEFE}
\end{eqnarray}
where $X_{\mu\nu}$ is the effective energy-momentum tensor obtained by varying the potential term with respect to $g_{\mu\nu}$,
\begin{eqnarray}
 X_{\mu\nu} &=& {\cal K}_ {\mu\nu} -{\cal K}g_ {\mu\nu} -\alpha\left\{{\cal K}^2_{\mu\nu}-{\cal K}{\cal K}_{\mu\nu} +\frac{[{\cal K}]^2-[{\cal K}^2]}{2}g_{\mu\nu}\right\} \nonumber\\
  && +3\beta\left\{ {\cal K}^3_{\mu\nu} -{\cal K}{\cal K}^2_{\mu\nu} +\frac{1}{2}{\cal K}_{\mu\nu}\left\{[{\cal K}]^2 -[{\cal K}^2]\right\} \right.\nonumber
  \\
 && \left. - \frac{1}{6}g_{\mu\nu}\left\{[{\cal K}]^3 -3[{\cal K}][{\cal K}^2] +2[{\cal K}^3]\right\} \right\} . \label{effemt}
\end{eqnarray}
In addition to the modified Einstein equations, one can obtain  {a} constraint by using the Bianchi identities as follows
\begin{eqnarray}\label{BiEoM}
 \nabla^\mu X_{\mu\nu} = 0,
\end{eqnarray}
where $\nabla^\mu$ denotes the covariant derivative which is compatible with $g_{\mu\nu}$.
Henceforth, we shall use $\alpha$ and
$\beta$, instead of the  parameters $\alpha_3$ and $\alpha_4$.

\section{Black hole solution in dRGT massive gravity} \label{solutions}
In this section, we will look for a static and spherically symmetric black hole solution of the modified Einstein equations  (\ref{EoM})  with the physical metric ansatz
\begin{eqnarray}\label{metric-gen}
 ds^2 = -n(r)dt^2 + 2d(r) dt dr +\frac{dr^2}{f(r)} + h(r)^2 d\Omega^2,
\end{eqnarray}
 The solution is found and classified into two branches: $d(r) = 0$ or $h(r) = h_0 r$ where $h_0$ is a constant in terms of the parameters $\alpha$ and $\beta$ \cite{Koyama:2011yg,Koyama:2011xz,Sbisa:2012zk}.   The most interesting branch is the diagonal branch, $d(r) = 0$, since it is simpler to analyze. For example, a class of a charged black holes within the diagonal branch in dRGT massive gravity was also investigated in Ref. \cite{Cai:2012db}.

The exact solutions for this ansatz are complicated and thus it is difficult to use these solutions to analyze the properties of black hole. Note that one may simplify the solution by choosing some specific relations of the parameters $\alpha$ and $\beta$ for example $\alpha = - 3 \beta$ \cite{Berezhiani:2011mt}. Furthermore, a class of parameters satisfying $\beta=\frac{\alpha^2}{3}$ simply yields the Schwarzschild-de Sitter solution \cite{Kodama:2013rea} (see also the analyses of such solution in Ref. \cite{Arraut:2014sja,Arraut:2014iba,Arraut:2014uza,Arraut:2015dva}).

 It is important to note that most solutions are asymptotically de Sitter or anti-de Sitter. This is not surprising since at large scale the theory should recover the  {cosmological} solution in which the graviton mass will play the role of cosmological constant to drive the late-time acceleration of the Universe. However, a class of the black hole solution in dRGT massive gravity (or in other classes of massive gravity, e.g. model in Ref. \cite{Dubovsky:2004sg})  may encounter the issues of superluminality, the Cauchy problem, and strong coupling. (see \cite{Motloch:2015gta} for the issues in dRGT and also \cite{Capela:2011mh,Deser:2013eua,Deser:2013qza,Izumi:2013poa} for those in another model of massive gravity)

Since the fiducial metric seems to play the role of a Lagrange multiplier to eliminate the BD ghost, one can choose an appropriate form to simplify the calculation. In the present work, we will follow \cite{Vegh:2013sk,Cai:2014znn,Adams:2014vza,Xu:2015rfa} by choosing the fiducial metric  to be
\begin{eqnarray}\label{fiducial metric}
f_{\mu\nu}=\text{diag}(0,0,c^2  ,c^2 \sin^2\theta), \label{fmetric}
\end{eqnarray}
where $c$ is a constant.  With the choice of the fiducial metric, the
action remains finite since it only contains non-negative powers of
$f_{\mu\nu}$ (see \cite{Vegh:2013sk}, for more details). It is important to note that the effective energy momentum tensor in Eq. (\ref{effemt}) is derived by requiring that the fiducial metric must be non-degenerate. From Eq. (\ref{fiducial metric}), it is obvious that the fiducial metric is degenerate and then one may not use the effective energy momentum tensor expressed in Eq. (\ref{effemt}). However, as point out in \cite{Cao:2015cti}, one can use the Moore-Penrose pseudoinverse of the metric ${\cal K}^\mu_\nu$ in order to find the effective energy momentum tensor and it provides the same expression in Eq. (\ref{effemt}). Therefore, one can use the effective energy momentum tensor defined in Eq. (\ref{effemt}) for the form of the fiducial metric. Moreover, the results can be checked by using the mini-superspace   action as usually done in cosmological analysis. One can obtain the equation of motion by using the Euler-Lagrangian equation for the variables, $n$ and $f$. As a result, we found that the equations of motion still valid.

For the physical metric, we will consider the diagonal branch of the physical metric by setting $d(r)=0$. In order to obtain  {a} black hole solution, we will choose the function $h(r) = r$.  {Then,} the physical metric can be written as
\begin{eqnarray}\label{metric}
 ds^2 = -n(r)dt^2 +\frac{dr^2}{f(r)} +r^2 d\Omega^2. \label{metric}
\end{eqnarray}
Up to this point, we have chosen to investigate just a class of the solutions to the dRGT massive gravity which possesses symmetries of our interest. Symmetries of solutions are of great important in massive gravity since they affect the number of degrees of freedom and the stability of the theory at times, which are also significant issues for many massive gravity models. 
As found in cosmological background, the number of degree of freedom crucially depends on the 
isotropy and homogeneity 
of the background physical metric as well as the form of the fiducial metric \cite{Gumrukcuoglu:2011ew,Gumrukcuoglu:2011zh,Chullaphan:2015ija}. Therefore, it is worthwhile to find, for this choice of the physical and the fiducial metric,whether the number of degree of freedom still valids and each of them is healthy. This issue is out of scope of this work and we leave this investigation for further work.

From the ansatz in Eq. (\ref{metric}), components of the Einstein tensor can be written as
\begin{eqnarray}
G^{t}_{t} &=& \frac{f'}{r}+\frac{f}{r^2}-\frac{1}{r^2},\\
G^{r}_{r} &=& \frac{f \left(r n'+n\right)}{n r^2}-\frac{1}{r^2},\\
G^{\theta}_{\theta} = G^{\phi}_{\phi} &=& f' \left(\frac{n'}{4 n}+\frac{1}{2 r}\right)+f \left(\frac{n''}{2 n}+\frac{n'}{2 n r}-\frac{\left(n'\right)^2}{4 n^2}\right). \label{EFE}
\end{eqnarray}
Computing the effective energy-momentum tensor in Eq. (\ref{effemt}) with this ansatz, the tensor $X_{\mu\nu}$ can be written as
\begin{eqnarray}
X^{t}_{t} &=& -\left(\frac{\alpha  (3 r-c) (r-c)}{r^2}+\frac{3 \beta  (r-c)^2}{r^2}+\frac{3 r-2 c}{r}\right), \label{Xtt}\\
X^{r}_{r} &=&-\left(\frac{\alpha  (3 r-c) (r-c)}{r^2}+\frac{3 \beta  (r-c)^2}{r^2}+\frac{3 r-2 c}{r}\right), \label{Xrr}\\
X^{\theta}_{\theta} = X^{\phi}_{\phi} &=& \frac{\alpha  (2 c-3 r)}{r}+\frac{3 \beta  (c-r)}{r}+\frac{c-3 r}{r}. \label{effemt2}
\end{eqnarray}
Note that $X^t_t=X^r_r$. There are specific values of the parameter $c$ by which this effective energy-momentum tensor behaves like a cosmological constant. In particular, $c=0$ simplifies each of the diagonal component of the effective energy-momentum tensor so that  {they depend} only on the parameters of the theory, which are all equal constants. Actually, by setting $c=0$ in Eq. (\ref{K-tensor}),  the tensor $\mathcal{K}^\mu_\nu$  will be equal to the identity matrix leading to the fact that the mass terms are all constants at the Lagrangian level  which is corresponding to the cosmological constant term. This is a crucially different point between our model and one in Ref. \cite{Cai:2014znn}. In our model, the solution can be reduced to Schwarzschild-AdS/dS solution where the cosmological constant-like term can be expressed in terms of the graviton mass while the cosmological constant-like term in Ref. \cite{Cai:2014znn} is introduced by hand and not related to the graviton mass.

Substituting all components of Einstein tensor and effective energy momentum tensor into Eq. (\ref{modEFE}), the modified Einstein equations can be written explicitly as
\begin{align}
\frac{f'}{r}+\frac{f}{r^2}-\frac{1}{r^2} &= m_{g}^2\left(\frac{\alpha  (3 r-c) (r-c)}{r^2}+\frac{3 \beta  (r-c)^2}{r^2}+\frac{3 r-2 c}{r}\right),\label{eom1}
\\
\frac{f \left(r n'+n\right)}{n r^2}-\frac{1}{r^2} &= m_{g}^2\left(\frac{\alpha  (3 r-c) (r-c)}{r^2}+\frac{3 \beta  (r-c)^2}{r^2}+\frac{3 r-2 c}{r}\right),\label{eom2}
\\
f' \left(\frac{n'}{4 n}+\frac{1}{2 r}\right)&+f \left(\frac{n''}{2 n}+\frac{n'}{2 n r}-\frac{\left(n'\right)^2}{4 n^2}\right) \nonumber
\\
&=  -m_{g}^2\left(\frac{\alpha  (2 c-3 r)}{r}+\frac{3 \beta  (c-r)}{r}+\frac{c-3 r}{r}\right).\label{eom3}
\end{align}
From Eq. (\ref{eom1}), one can obtain the solution of $f$  {which} can be written as
\begin{eqnarray}
f(r) =1 - \frac{2 M}{r}+ \frac{\Lambda}{3}r^2+\gamma r+\zeta, \label{solutionf}
\end{eqnarray}
where
\begin{subequations}
\begin{eqnarray}
\Lambda&=&3 m^2_g \left(1+\alpha+\beta\right),\\
\gamma&=&-c m^2_g \left(1+2\alpha+3\beta\right),\\
\zeta&=&c^2 m^2_g \left(\alpha+3\beta\right),
\end{eqnarray} \label{pardef}
\end{subequations}
and $M$ is an integration constant  related to mass of the black hole. This solution incorporates the cosmological constant term, namely $\Lambda$, naturally in terms of the graviton mass $m_g$ which should not be surprising since the graviton mass serves as the cosmological constant in the self-expanding cosmological solution in massive gravity. Moreover, this solution can be identified with the known solutions in general relativity. Note that the solutions in alternative forms with other sets of parameters are shown explicitly in \ref{A.A}. In the case $m_g=0$ we have the Schwarzschild solution, as expected. For $c=0$ which sets $\gamma=\zeta=0$, the solution can be classified according to the values of $\alpha$ and $\beta$. If $\left(1+\alpha+\beta\right)<0$, the solution is in the form of Schwarzschild-de-Sitter while the case $\left(1+\alpha+\beta\right)>0$ yields the Schwarzschild-Anti-de Sitter solution. Using Eq. (\ref{eom1}) and Eq. (\ref{eom2}), one finds that
\begin{eqnarray}
 n' f = f' n. \label{constraint}
\end{eqnarray}
This equation implies that function $f$ and $n$ differ only by a constant. One can choose the constant to obtain a black hole solution such that
\begin{eqnarray}
n(r)=f(r)=1 - \frac{2 M}{r}+ \frac{\Lambda}{3}r^2+\gamma r+\zeta. \label{nfsolution}
\end{eqnarray}
It may be verified by direct substitution this solution into Eq. (\ref{eom3}).  The dRGT
solution outlined here contains, for instance, the Schwarzschild  {solution} ($m_g=0)$,
the de Sitter/Anti-de Sitter  solutions, the global monopole solution of general relativity, and thus also contains the monopole-de Sitter-Schwarzschild solution. Note that the last term, the constant potential $\zeta$, corresponds to the global monopole term. A global
monopole solution was introduced by
Barriola and Vilenkin \cite{bv}, and usually comes from a topological defect in high energy physics at early universe resulting from a gauge-symmetry breaking \cite{Huang:2014oga,Tamaki:2003kv}. However, in this solution, the global monopole is contributed from a graviton mass.  In addition, we see that the solution here is similar to the four-dimensional solution found in Ref. \cite{Cai:2014znn}.  In Ref. \cite{Cai:2014znn}, the author obtained the black hole solution where the bare cosmological constant is presented while, on the other hand, our result incorporates the cosmological-constant-like behavior, namely the $\Lambda r^2$, in the black hole solution automatically due to introducing the potential term.
\begin{figure}[h!]
\includegraphics[scale=0.60]{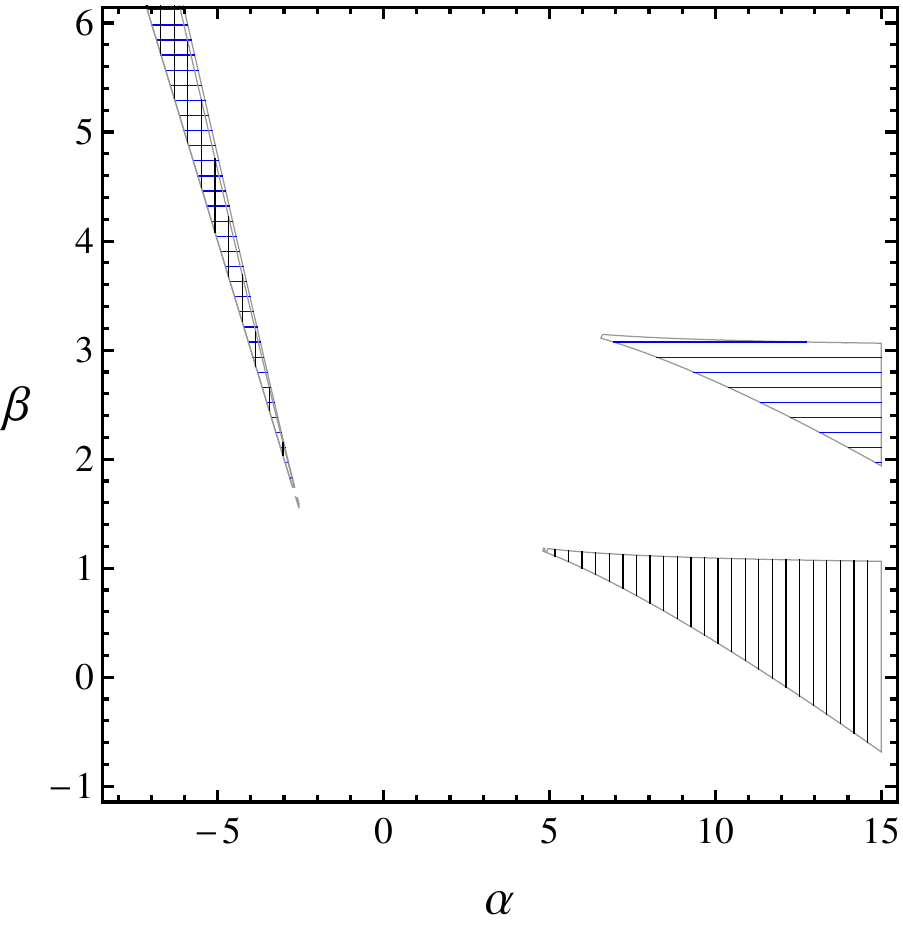}\qquad
\includegraphics[scale=0.63]{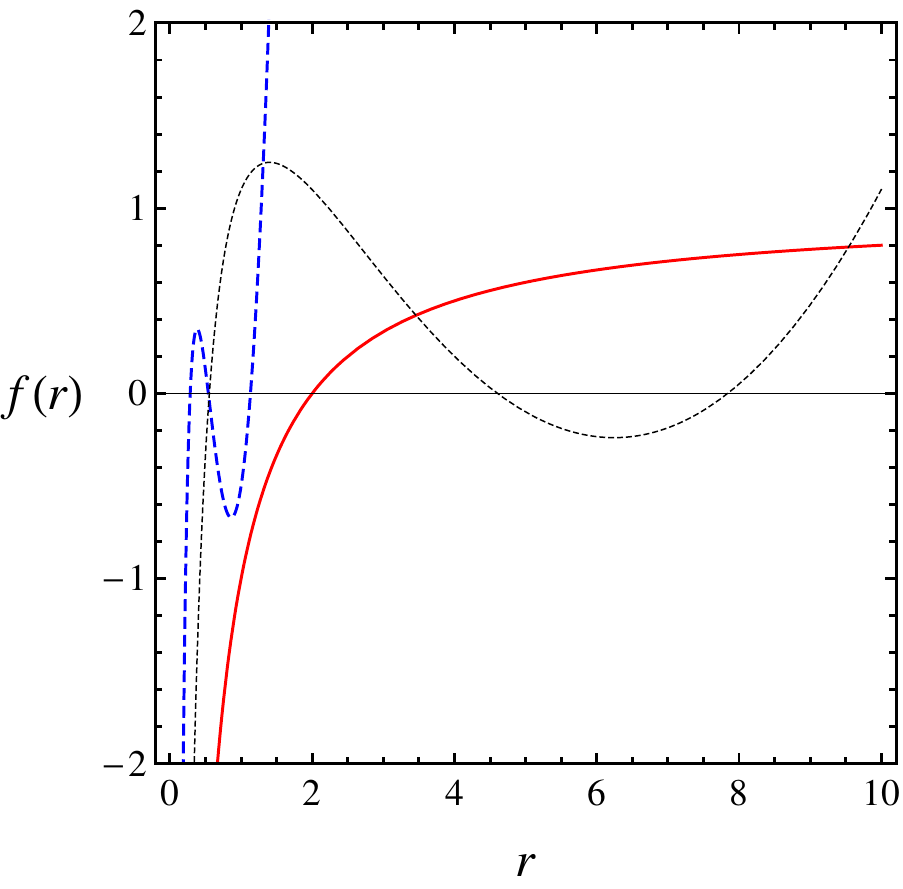}
{\caption{\label{BHhorizon}  The graph in the left panel shows the domain region in $(\alpha,\beta)$-space for the black hole solution with three event horizons. The horizontal-blue shaded region corresponds to the region with $m_g=c=1$ and $M=2$ while the vertical-black shaded region corresponds to the region with $m_g=c=1$ and $M=1$. Note that there is an overlapping region where the black hole can have three event horizons for both values of the mass $M$. For the right panel, the graph shows the profile of $f(r)$ (with $M=1$). The red-solid curve corresponds to the case $m_g=0$, which recovers the Schwarzschild geometry while the blue-dashed curve corresponds to a set of parameters $m_g=c=1$, $\alpha=-3$, and $\beta=2.1$. The black-dotted curve corresponds to set of parameters as $m_g= c=1$, $\alpha=10$, $\beta=0.5$. Together, these demonstrate the existence of three event horizons.}}
\end{figure}

However, this solution highly depends on the choice of the fiducial metric; changing to other forms of the fiducial metric will significantly affect the solution. This kind of dependency is one of the important properties of massive gravity. For example, from a cosmological point of view, one cannot have a nontrivial flat cosmological solution with a Minkowski fiducial metric \cite{D'Amico:2011jj}; only the open FLRW solution is allowed \cite{Gumrukcuoglu:2011ew}, where the FLRW solution with arbitrary geometry exists when the FLRW fiducial metric is considered \cite{Gumrukcuoglu:2011zh}. By generalizing the form of the fiducial metric, the nontrivial cosmological solutions can be obtained \cite{Chullaphan:2015ija}.

\section{Thermodynamics of the black hole}\label{thermo}
For black holes in de Sitter space, there exists more than one horizon, and the multiple horizons correspond
to different thermodynamic systems.  Next, we shall explore the thermodynamics of the dRGT massive gravity black hole solution given by the Eq. (\ref{solutionf})  by assuming that the black hole is a closed system, i.e.,  no particle (or charge) transfer or creation/annihilation.  {In other words,} the black hole is assumed to be a canonical ensemble system.  The horizons, if they exist, are given by zeros of $g^{rr} = 0$ \cite{Ghosh:2014pga} or
\begin{eqnarray}
1 - \frac{2 M}{r}+ \frac{\Lambda}{3}r^2+\gamma r+\zeta = 0, \label{horizoneq}
\end{eqnarray}
which may admit three real roots. For some classes of parameter setup, there may exist up to three horizons as shown Figure \ref{BHhorizon}. From the left panel in this figure, we use a region plot to find the region in $(\alpha,\beta)$-space for the existence of three positive real roots of Eq. (\ref{horizoneq}) by setting $m_g = 1$, $c=1$. We also pick up two points in this region to show the profile of $f(r)$ in the right panel of this figure.  We note that the gravitational mass of a black hole is determined by $f(r_+) = 0$, which in terms of the outer horizon radius $r_+$ reads
\begin{eqnarray}
M &=&\frac{r_+}{2}\left(1+ \frac{\Lambda}{3}r^2_+ +\gamma r_+  +\zeta\right), \label{bhmass}\\
\frac{M}{c} &=& \frac{\bar{r}_+}{2}\left(1+ m^2_g c^2 \left[(1+\alpha+\beta)\bar{r}_+^2 -(1+2\alpha+3\beta)\bar{r}_+ + (\alpha + 3\beta)\right]\right),\label{bhmass-ab}
\end{eqnarray}
where $\bar{r}_+ = r_+/c$. For simplicity, one can work in dimensionless parameters with units of $m^2_g c^2 = 1$. In this unit, the conditions for the positive value of the black hole mass can be written as
\begin{eqnarray}
\alpha &>& -\frac{(1+\beta)\bar{r}_+^2 -(1+3\beta)\bar{r}_+ +(1+3\beta)}{(\bar{r}_+-1)^2}\,\,\text{for}\,\,\, \bar{r}_+\neq 1,\\
\beta &>&-1\,\, \text{and}\,\, \alpha\,\, \text{is arbitrary}\,\, \text{for}\,\,\bar{r}_+ = 1.
\end{eqnarray}
The Hawking temperature associated with the black hole is related with the surface gravity ($\kappa$) via $T=\kappa/(2 \pi)$.
The surface gravity in terms of the metric function  reads $\kappa=\frac{f'(r_+)}{2}$ \cite{Ghosh:2014dqa} and hence the temperature $T$ of the black hole becomes
\begin{eqnarray}
T=\frac{1}{4\pi r_+}\left(1+\Lambda r^2_+ +2\gamma r_+ +\zeta \right). \label{bhtemp}
\end{eqnarray}
Taking the limit $m_g=0$, one recovers the temperature for the Schwarzschild black hole \cite{Ghosh:2014dqa}:
\begin{eqnarray}
T=\frac{1}{4\pi r_+}
\end{eqnarray}
\begin{figure}[h!]
\includegraphics[scale=0.54]{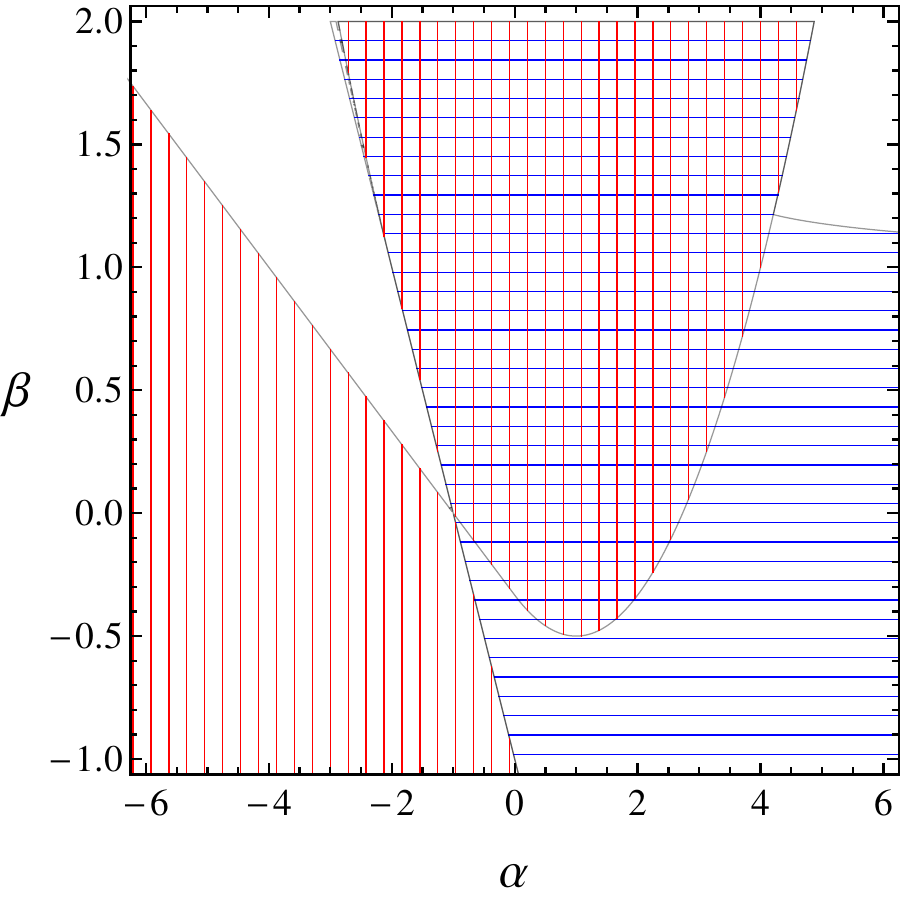}\qquad\qquad
\includegraphics[scale=0.54]{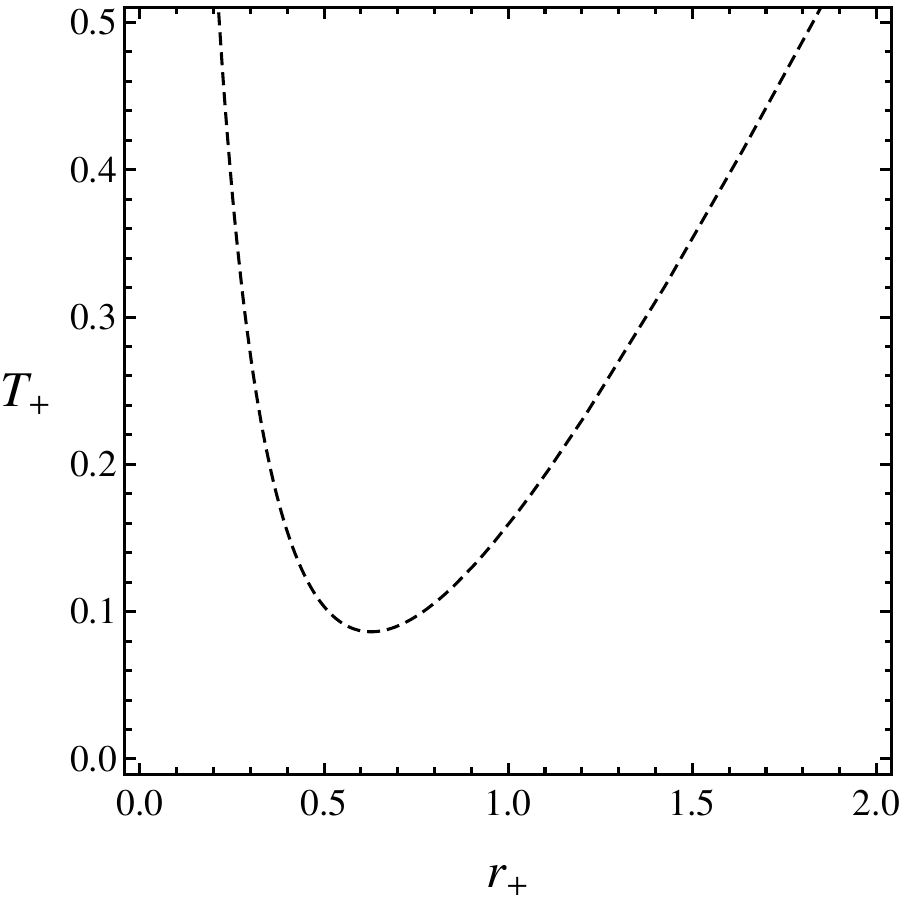}
{\caption{In the left panel, the graph shows the domain regions in $(\alpha,\beta)$-space for the condition $T_{+(\text{min})} > 0$ corresponding to the vertically-red-striped region and for the condition $r_+ > 0$ corresponding to the horizontally-blue-striped region. The intersection corresponds to the region satisfying both conditions; $T_{+(\text{min})} > 0$ and $r_+ > 0$.  We choose the point in the intersection region as $ \alpha=1, \beta=0.2$ to illustrate the temperature profile of the black hole given by Eq. (\ref{bhtemp}) in the right panel. Note that we have used the unit of $m^2_g c^2 =1$ for both plots and also set $M=1$ for the plot in the left panel.  }\label{BHtempfig}}
\end{figure}
The crucially different point in the temperature profile compared with one obtained in the Schwarzschild solution is that it is possible to find the positive local minimum of the temperature as shown in the right panel of Figure \ref{BHtempfig}. In the left panel of this figure, we use units of $m^2_g c^2 = 1$ and show the region matching the requirement of positivity of the local minimum of the temperature, $T_{+(\text{min})} > 0$, together with the positivity of the horizon size, $r_+ > 0$. We also adopt a simple choice of the parameters in this region, such as $(\alpha=1, \beta=0.2)$, to illustrate the existence of the positive local minimum of the temperature as shown in the right panel of Figure \ref{BHtempfig}. Note that, in the unit of $m^2_g c^2 = 1$, the dimensionless version of the temperature and horizon size can be written as $\bar{T} = T c$ and $\bar{r}_+ = r_+ / c$. These are the actual values of the quantities plotted in the right panel of Figure \ref{BHtempfig}. For simplicity, we set $c= 1$ which leads to $m_g = 1$ and $\bar{T} = T $ as well as $\bar{r}_+ = r_+ $.

Next, we turn our attention to the important thermodynamic quantity
associated with the black hole horizon which is its entropy ($S$). The black hole is supposed to obey the first law of thermodynamics,  {or} $dM = TdS$. To calculate the entropy, we use
\begin{equation}\label{intS}
S = \int T^{-1} dM =\int T^{-1} \frac{\partial M}{\partial r+} dr_+.
\end{equation}
Integrating the above equation leads to
\begin{eqnarray}
S=\pi r^2_+. \label{bhentropy}
\end{eqnarray}
This can be written as $S=\frac{A_+}{4}$ with the area of the horizon $A_{+} =4\pi r_+^2$, which is the famous area law.  Interestingly,  {the graviton mass does not significantly affect the form of the entropy; it contributes only as a correction for the horizon radius which can be seen explicitly from Eq. (\ref{bhmass-ab}).}
The behavior of the temperature and its corresponding horizon is shown in Figure \ref{BHtempfig},  {in} which there exists a local minimum of the temperature. This feature suggests that there should be a \emph{transition} between two states; from the \emph{non-black hole} or \emph{hot flat space} state to a black hole. The transition was realized by Hawking and Page \cite{Hawking:1982dh} (see also \cite{York:1986it}), who found the characteristics of the so-called Hawking-Page phase transition.
The transition exists if it is thermodynamically spontaneous, or alternatively, globally thermodynamically stable, by evaluating the free energies between two states. In other words, this corresponds to evaluating the Euclidean actions of those two states where the temperature is treated as a period of the imaginary time. Since in this case, there is no particle transfer, we compute the Helmholtz free energy
\begin{eqnarray}
F &=& M-TS, \nonumber
\\
&=& \frac{r_+}{2}\left(1+ \frac{\Lambda}{3}r^2_+ +\gamma r_+  +\zeta\right)-\frac{r_+}{4}\left(1+\Lambda r^2_+ +2\gamma r_+ +\zeta \right) \nonumber
\\
&=& \frac{r_+}{4}\left(1-\frac{\Lambda}{3}r^2_+ +\zeta\right).
\end{eqnarray}
The existence of globally thermodynamical stability of the black hole is determined by the condition $F \leq 0$. Therefore, the transition exists if
\begin{eqnarray}
\Lambda r^2_+ \geq 3\left(1+\zeta\right), \label{bhHPcon}
\end{eqnarray}
where the transition from the vacuum state to the black hole takes place at $F=0$.

\begin{figure}[h!]
\begin{center}
\includegraphics[scale=0.6]{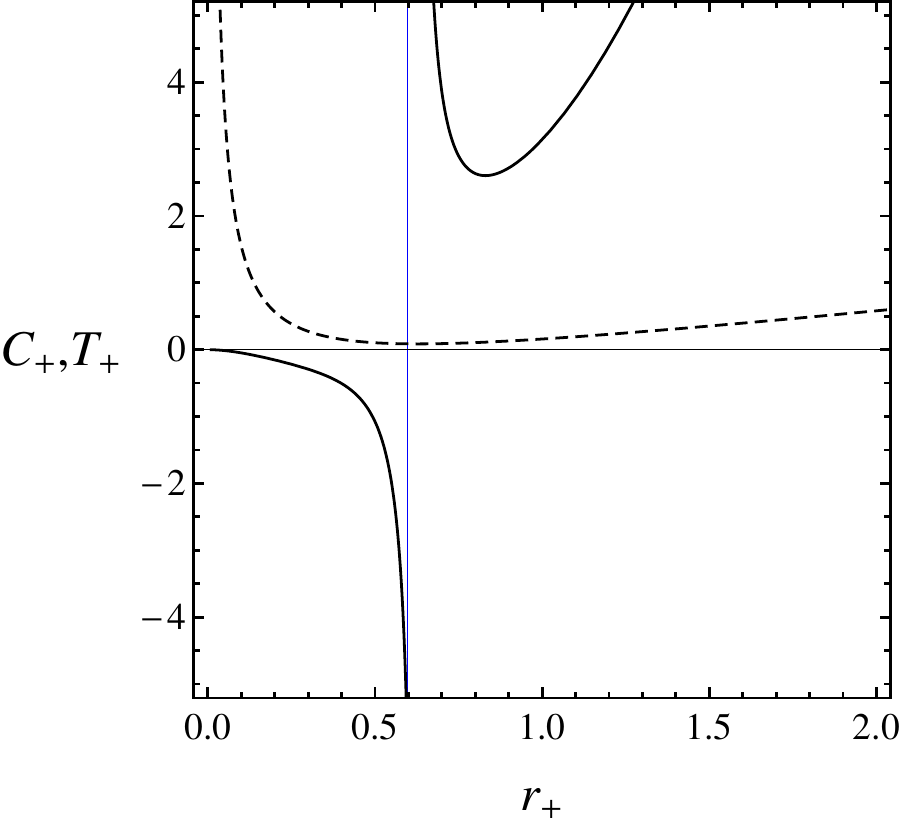}
\end{center}
{\caption{ {This is} the plot of the heat capacity of the black hole, corresponding to the black-solid line, given by Eq. (\ref{bhheatcap}) with the parameter  {setup} $m_g=c=1, \alpha=1, \beta=0.2$. The heat capacity diverges at the radius to which the minimal temperature is associated. The graph also shows the temperature profile in dashed-black curve.}\label{BHheatfig}}
\end{figure}

In addition to the globally thermodynamical stability, one can determine the locally thermodynamical stability by examining the sign of the heat capacity. The heat capacity of the black hole is given by
\begin{eqnarray}
C_+=\left(\frac{\partial M}{\partial T}\right)_{r=r_+} = \left(T\frac{\partial S}{\partial T}\right)_{r=r_+}.
\end{eqnarray}
By using the relations in Eq. (\ref{bhmass}) and Eq. (\ref{bhtemp}) , the heat capacity becomes
\begin{eqnarray}
C_+=\frac{2\pi r^2_+ \left(1+2\gamma r_+ +\Lambda r^2_+ +\zeta\right)}{\Lambda r^2_+ - \left(1+\zeta\right)}=\frac{8\pi^2 r^3_+ T}{\Lambda r^2_+ - \left(1+\zeta\right)}. \label{bhheatcap}
\end{eqnarray}

One can see from Figure {\ref{BHheatfig}}. that there is a particular horizon  for which the corresponding heat capacity diverges. The divergence is due to the minimum temperature and the corresponding horizon is exactly the horizon of minimum temperature. 
Furthermore, requiring the locally thermodynamical stability  of the black hole, the black hole must obey the condition
\begin{eqnarray}
\Lambda r^2_+ > \left(1+\zeta\right). \label{bhheatcapcon}
\end{eqnarray}
\begin{figure}[h!]
\includegraphics[scale=0.6]{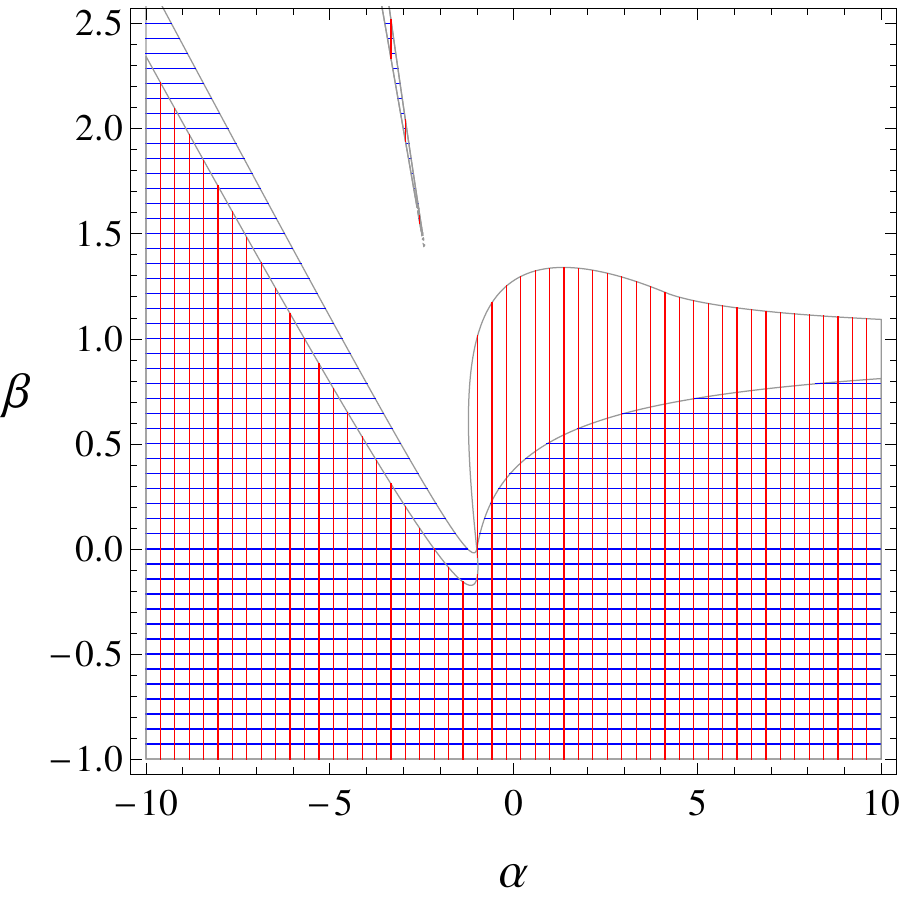}\qquad\quad
\includegraphics[scale=0.58]{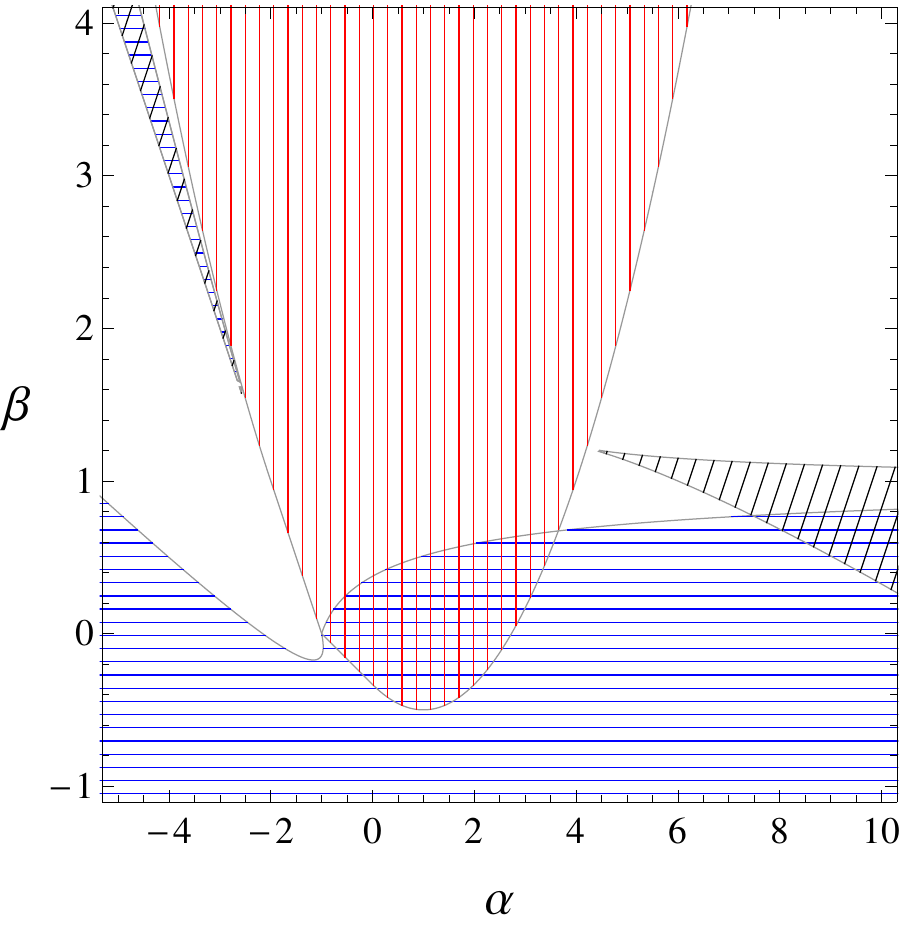}
{\caption{The left panel is the region plot of $(\alpha,\beta)$-space which satisfies the stability conditions with given values of the parameters; $m_g = 1$, $c=1$, and $M=1$. The blue-shaded horizontally-striped region corresponds to the values of $\alpha$ and $\beta$ by which the condition for globally thermodynamical stability in Eq.(\ref{bhHPcon}) is satisfied while the vertically-red-striped region implies locally thermodynamical stability according to the Eq. (\ref{bhheatcapcon}). The intersection corresponds to the region which satisfies both globally and locally thermodynamical stabilities. For the right panel, the graph shows the region satisfying both of the thermodynamical stability conditions in blue-shaded horizontally-striped region, the region satisfying the conditions $T_{+(\text{min})} > 0$ and $r_+ > 0$ in vertically-red-striped region, and the region in which three-event-horizon black hole exists in the skewed-black-striped region.   }\label{alphabetaplot}}
\end{figure}
One can see from Eq. (\ref{bhHPcon}) and Eq. (\ref{bhheatcapcon}) that both of the thermodynamical stabilities depend on the parameters $\Lambda$ and $\zeta$ which are determined by the parameters of the massive gravity theory, namely, $\alpha$ and $\beta$. We use the plot to find the allowed region in $(\alpha, \beta)$-space by using the unit of $m^2_gc^2=1$ and setting $M=1$. We show the validity of those parameters where both the stabilities are assumed in the left panel of Figure \ref{alphabetaplot}. Note that one can express $r_+$ in terms of $\alpha$, $\beta$ and $M$ from Eq. (\ref{bhmass-ab}) and then substitute this expression into the stability conditions to find stability regions in $(\alpha, \beta)$ space where $M$ is held fixed to be a positive constant. From this figure, one can see that there exists an allowed region for the transition phase with the parameters $\alpha, \beta \sim O(1)$. This suggests that massive gravity can naturally provide the Hawking-Page phase transition without requiring fine-tuning of the parameters. For the right panel of Figure \ref{alphabetaplot}, we combine three important regions according to the previous plots including the regions satisfying thermodynamical stability conditions (intersection region in the left panel of Figure \ref{alphabetaplot}), $T_{+(\text{min})} > 0$ and $r_+ > 0$ (intersection region in left panel of Figure \ref{BHtempfig}), and the existence of three horizons (region in the left panel of Figure \ref{BHhorizon}).  From this figure, one can see that the thermodynamically stable region of the black hole together with positive temperature and horizon size is not compatible with the region indicating the existence of three horizons. {This is due to the fact that the temperature is proportional to $f'(r_+)$ and the condition of existence of three horizons is that there exist two extremum points such that $f'(r) = 0$. This implies that there always exists a range of horizon ($r_+$) at which the black hole temperature is negative if there exist three real black hole horizons. } Therefore, we use a set of parameters which give rise to one or two horizons to illustrate the thermodynamical quantities such as temperature and heat capacity.

\begin{figure}[h!]
\includegraphics[scale=0.6]{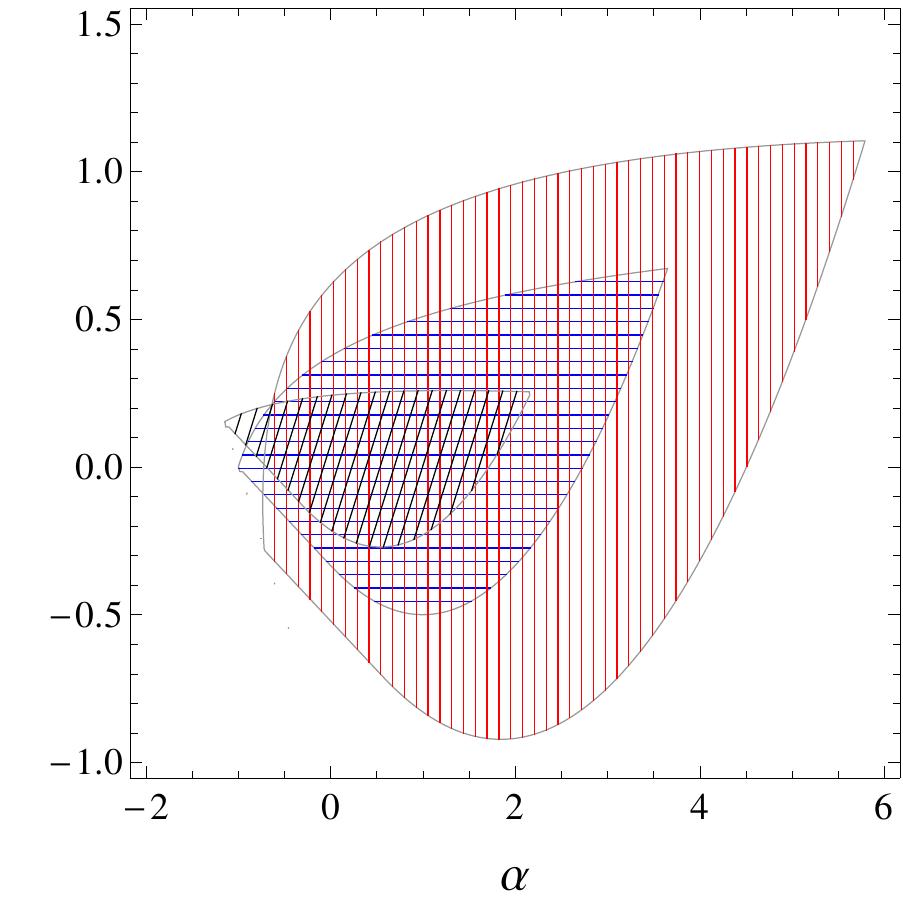}\qquad\qquad
\includegraphics[scale=0.56]{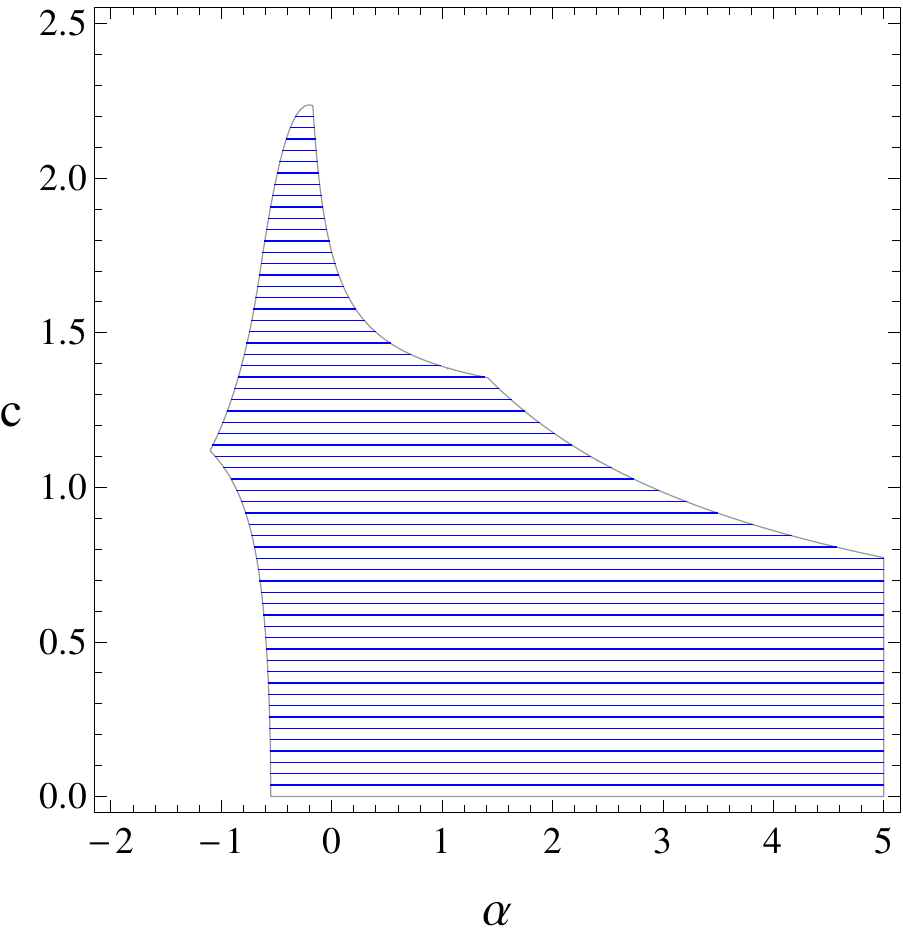}
{\caption{In the left panel, the graph shows the stability region together with positivity of $T_{+(\text{min})} > 0$ and $r_+ > 0$ with different values of the parameter $c$. The vertically-red-striped region, the horizontally-blue-striped region, and the skewed-black-striped region correspond to the region with $c= 0.8$,  $c = 1 $ and $c=1.2$ respectively. In the right panel, we choose a common value of parameter with $\beta = 0.1$ to illustrate the stability region in ($\alpha, c$) space. Both figures show that the greater the value of the parameter $c$, the smaller the region of stability. }\label{stab-c}}
\end{figure}
It is worthwhile to note that in the expression of dimensionless variables, such as Eq. (\ref{bhmass-ab}), the dimensionless variable of $r_+$ is $\bar{r}_+ = r_+/c$. Therefore, the dimensionless horizon size is inversely proportional to the parameter $c$. This means that the greater the value of the horizon size, the smaller the value of the parameter $c$. In the left panel of Figure \ref{stab-c}., we explicitly show the stability region in $(\alpha,\beta)$-space with different values of the parameter $c$ such that $c=0.8, c=1.0$ and $ c=1.2$. Furthermore, we also show the allowed region in $(\alpha, c)$-space by fixing $\beta = 0.1$ in the right panel. From this figure, it is found that the greater  value of the parameter $c$ corresponds to the smaller value of the horizon size and the smaller allowed region in the parameter space. Therefore, the phase transition tends to occur more easily at large horizon size. Note that in this case, the black hole is treated as a canonical ensemble system  {where} particle transfer is prohibited. We will discuss the charged black hole case in the next section.

\section{Charged black hole} \label{charge}
\begin{figure}[h!]
\includegraphics[scale=0.6]{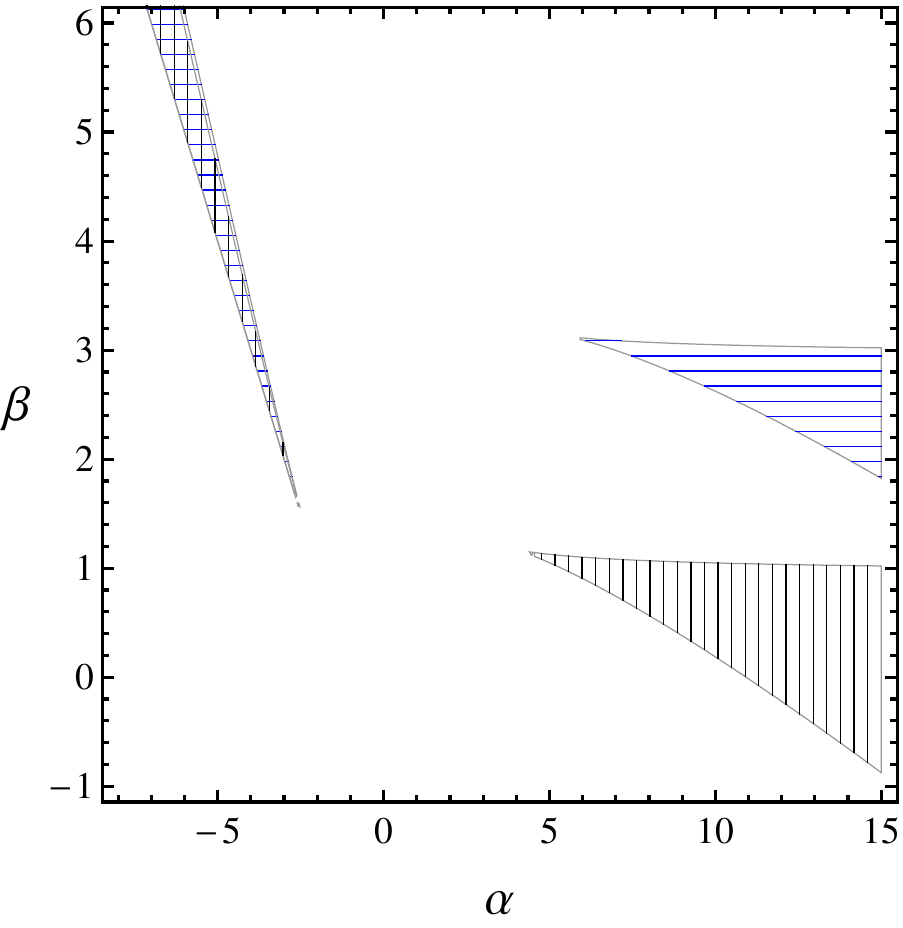}\qquad
\includegraphics[scale=0.64]{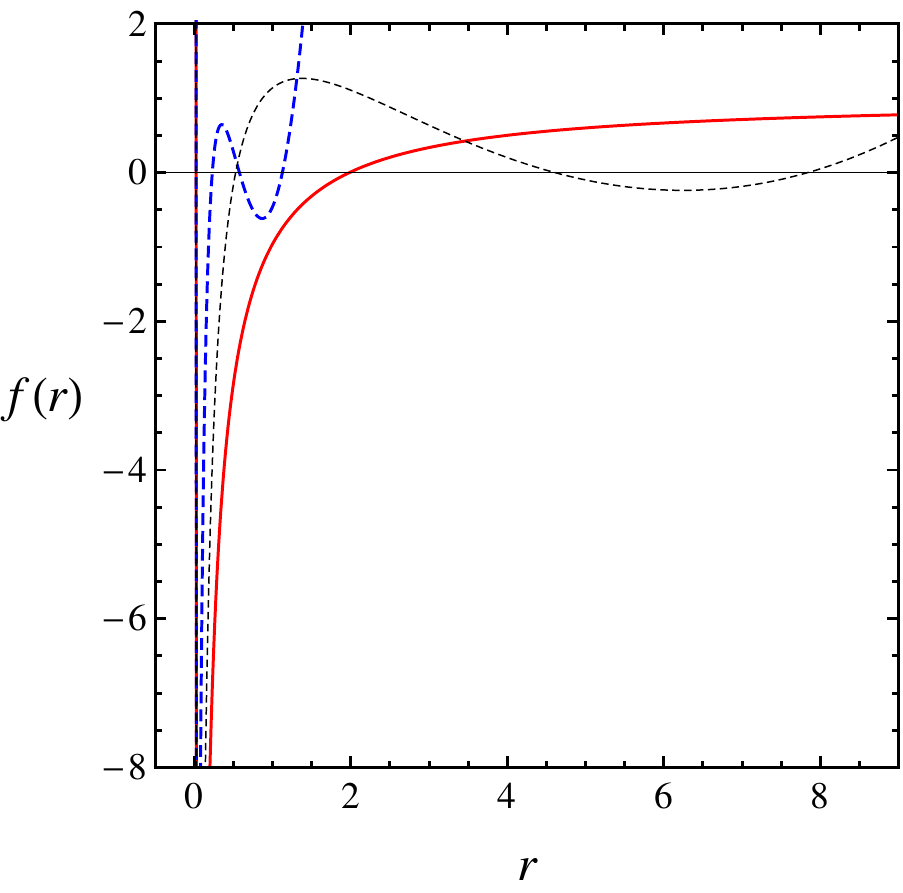}
{\caption{In the left panel, the graph shows the domain region in $(\alpha,\beta)$-space for the black hole having four event horizons. The horizontally-blue-striped region corresponds to the region with $m_g=1$, $c=1$, $Q=0.2$ and $M=2$ while the vertically-black-striped region corresponds to the region with $m_g=1$, $c=1$, $Q=0.2$ and $M=1$. For the right panel, the graph shows the profiles of $f(r)$ with charge (the blue-dashed curve and black-dotted curve) compared with the Reissner-Nordstr$\ddot{\text{o}}$m solution (the red solid curve). The blue-dashed and black-dotted curves show an example of the existence of four horizons for the charged black hole, with parameter set $m_g=1$, $Q=0.2$, $c=1$, $\alpha=-3$, $\beta=2.1$, and $m_g=1$,  $Q=0.2$, $c=1$, $\alpha=10$, $\beta=0.5$ respectively.
 }\label{BHChorizon}}
\end{figure}

In this section, the black hole with non-zero charge and both the grand canonical aspect of the black hole, where the charge transfer is allowed, and the canonical ensemble point of view will be discussed.
Hence, it will be interesting to consider the charged generalization of the above solution.  The action in Eq. (\ref{action}) with Maxwell term reads
\begin{eqnarray}
S = \int d^4x \sqrt{-g}  \Bigg[\frac{1}{2\kappa^2}\left( R +m_g^2 {\cal U}(g, \phi^a)\right)-\frac{1}{16\pi}F_{\mu\nu}F^{\mu\nu}\Bigg], \label{actioncharge}
\end{eqnarray}
where $F_{\mu\nu}\equiv\left(\nabla_\mu A_\nu-\nabla_\nu A_\mu\right)$ is the Maxwell strength tensor and $A_\mu$ is a vector potential (the action is written in the Gaussian unit). We limit our study by considering a spherically-symmetric dRGT black hole filled only with a static charge which is accompanied by $A_\mu = \left(A(r),0,0,0\right)$. The presence of the static charge gives rise to a non-vanishing energy-momentum tensor as
\begin{align}
T^\mu_{\nu} = \text{diag}\left(-\frac{fA^2}{8\pi n},-\frac{fA^2}{8\pi n},\frac{fA^2}{8\pi n},\frac{fA^2}{8\pi n}\right).
\end{align}
Thus, one can rewrite the Einstein equations in Eq. (\ref{eom1}) and Eq. (\ref{eom2}) with the presence of a static charge as
\begin{align}
\frac{f'}{r}+\frac{f}{r^2}-\frac{1}{r^2} &= -m_{g}^2X^t_t- \frac{fA^2}{n}, \label{eom1a}
\\
\frac{f \left(r n'+n\right)}{n r^2}-\frac{1}{r^2} &= -m_{g}^2X^r_r- \frac{fA^2}{n}, \label{eom2a}
\end{align}
where $X^t_t$ and $X^r_r$ are given in Eq. (\ref{Xtt}) and Eq. (\ref{Xrr}). Taking these equations into account, one can find that the constraint in Eq. (\ref{constraint}) still holds for the presence of a non-zero charge. This constraint also simplifies the equation of motion of $A(r)$.
By requiring the spherically symmetric solution as in Eq. (\ref{metric}), one find the equation of motion of $A(r)$ as
\begin{align}
A'' + A'\left(\frac{2}{r}+\frac{1}{2}\left(\frac{f'}{f}-\frac{n'}{n}\right)\right)&=0,
\\
A'' + A'\frac{2}{r}&=0,
\end{align}
where the constraint in Eq. (\ref{constraint}) is applied in the calculation. This equation simply implies the solution of the form
\begin{align}
A(r) = \frac{k}{r} + V_0,
\end{align}
which is exactly an electrostatic potential in a generic electrodynamics, where $k$ is an integration constant. The corresponding electric field is
\begin{align}
E(r) = - \vec{\nabla} A(r) = \frac{k}{r^2}.
\end{align}
To determine the value of $k$, one may consider an electric field from a charge $Q$ at large $r$ where the spacetime is asymtotically flat which, in the Gaussian unit, must take the form,
\begin{align}
E(r\rightarrow\infty) \sim \frac{Q}{r^2},
\end{align}
Obviously, this  implies that the integration constant $k$ must be identified to the charge $Q$. By solving Eq. (\ref{eom1a}) and Eq. (\ref{eom2a}), one finds a charged dRGT black hole solution as
\begin{eqnarray}
f^{(Q)}(r)=1 - \frac{2 M}{r}+\frac{Q^2}{r^2} + \frac{\Lambda}{3}r^2+\gamma r+\zeta, \label{solutionfcharge}
\end{eqnarray}
where $\Lambda, \gamma, \zeta$ are defined similarly as in Eq. ({\ref{pardef}}), note that we still have the relation in Eq. (\ref{nfsolution}). In general, this solution  {may have} up to 4 horizons, as illustrated in Figure \ref{BHChorizon}. Again, the innermost 2 horizons are the horizons that can be found in the Reissner-Nordstr$\ddot{\text{o}}$m black hole in the general relativity theory, where the others are the cosmological horizons associated with the existence of the graviton mass.

Following the same procedures as in section \ref{thermo}, one can find the mass and the temperature evaluated at the black hole horizon as
\begin{eqnarray}
M&=&\frac{r_+}{2}\left(1+\frac{Q^2}{r^2_+} + \frac{\Lambda}{2}r^2_+ +\gamma r_+ +\zeta\right), \label{bhcmass}
\\
T&=&\frac{1}{4\pi r_+}\left(1-\frac{Q^2}{r^2_+}+\Lambda r^2_+ +2\gamma r_+ +\zeta \right). \label{bhctemp}
\end{eqnarray}
Since there is non-zero charge involved, the thermodynamics will be different. In detail, one can treat the black hole as an open system, i.e. a grand canonical ensemble, where the charge transfer is allowed while another can view the system as a closed one or a canonical ensemble where the charge is a non-zero constant.

\subsection{Grand Canonical Ensemble}
To treat it as a thermodynamical object, one can consider the black hole as a grand canonical ensemble system where the chemical potential is held fixed as $\mu = \frac{Q}{r_+}$ (in the Gaussian unit). The corresponding temperature and entropy are
\begin{eqnarray}
T_g&=&\frac{1}{4\pi r_+}\left(1+\Lambda r^2_+ +2\gamma r_+ +\zeta -\mu^2\right),\label{bhcgtemp}
\\
S&=&\pi r^2_+. \label{bhcgentropy}
\end{eqnarray}
From Eq. (\ref{bhcgtemp}), the temperature will be lower due to the effect of the chemical potential, $\mu$.  In the grand canonical ensemble, the corresponding free energy is given as the Gibbs free energy,
\begin{eqnarray}
G&=&M-TS-\mu Q, \nonumber
\\
&=&\frac{1}{4}r_+ \left(1-\frac{\Lambda}{3}r^2_+ +\zeta -\mu^2\right). \label{Gibb-FE}
\end{eqnarray}
From this expression, one can see that the free energy shift to the smaller value due to the contribution from the chemical potential in the last term. In other words, the contribution from charge makes the free energy more negative. The sign of the free energy depends on the value of $\left(1-\frac{\Lambda}{3}r^2_+ +\zeta -\mu^2\right)$, in other words, there exists globally thermodynamical stability when
\begin{eqnarray}
\Lambda r^2_+ \geq 3\left(1 +\zeta -\mu^2\right) \label{bhcgHPcon}
\end{eqnarray}
 is satisfied. Moreover, one can check  locally thermodynamical stability by considering the heat capacity,
\begin{eqnarray}
C_g&=&\frac{2\pi r^2_+ \left[\left(1+2\gamma r_+ +\Lambda r^2_+ +\zeta\right)-\mu^2\right]}{\Lambda r^2_+ - \left(1+\zeta\right)+\mu^2} = \frac{8\pi^2 r^3_+ T}{\Lambda r^2_+ - \left(1+\zeta\right)+\mu^2}. \label{bhcgheatcap}
\end{eqnarray}
Requiring {locally} thermodynamical stability, one should have
\begin{eqnarray}
\Lambda r^2_+ > \left(1+\zeta - \mu^2\right). \label{bhcgheatcapcon}
\end{eqnarray}

\begin{figure}[h!]
\includegraphics[scale=0.62]{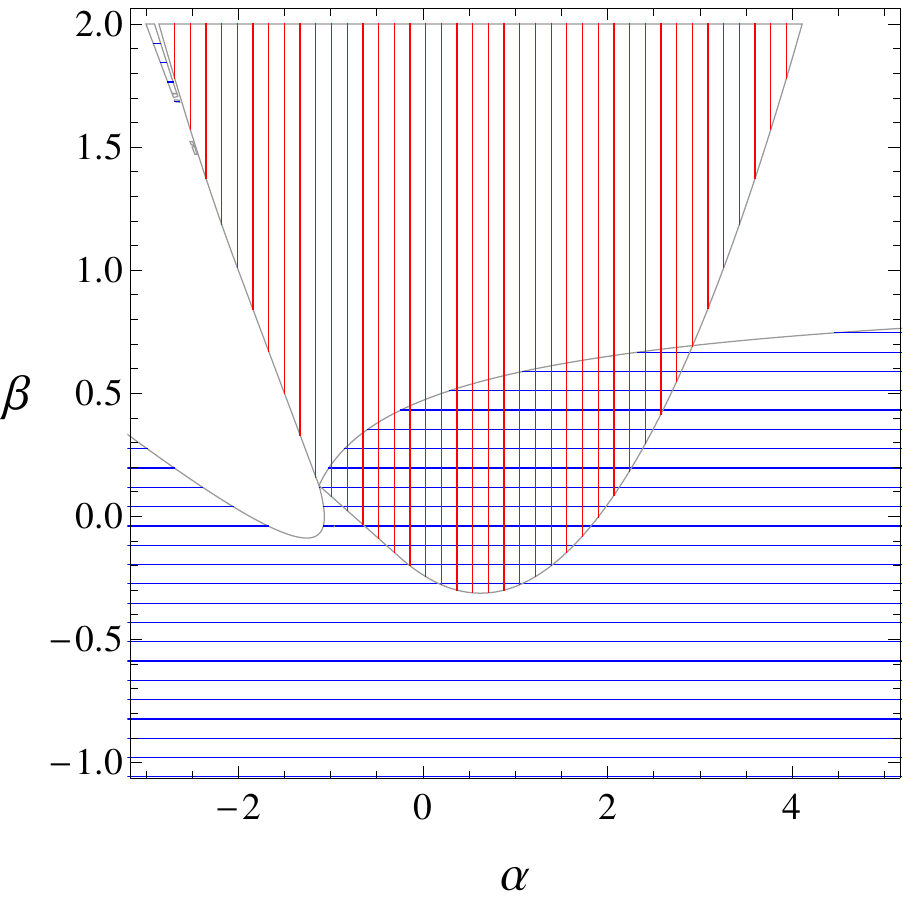}\qquad
\includegraphics[scale=0.62]{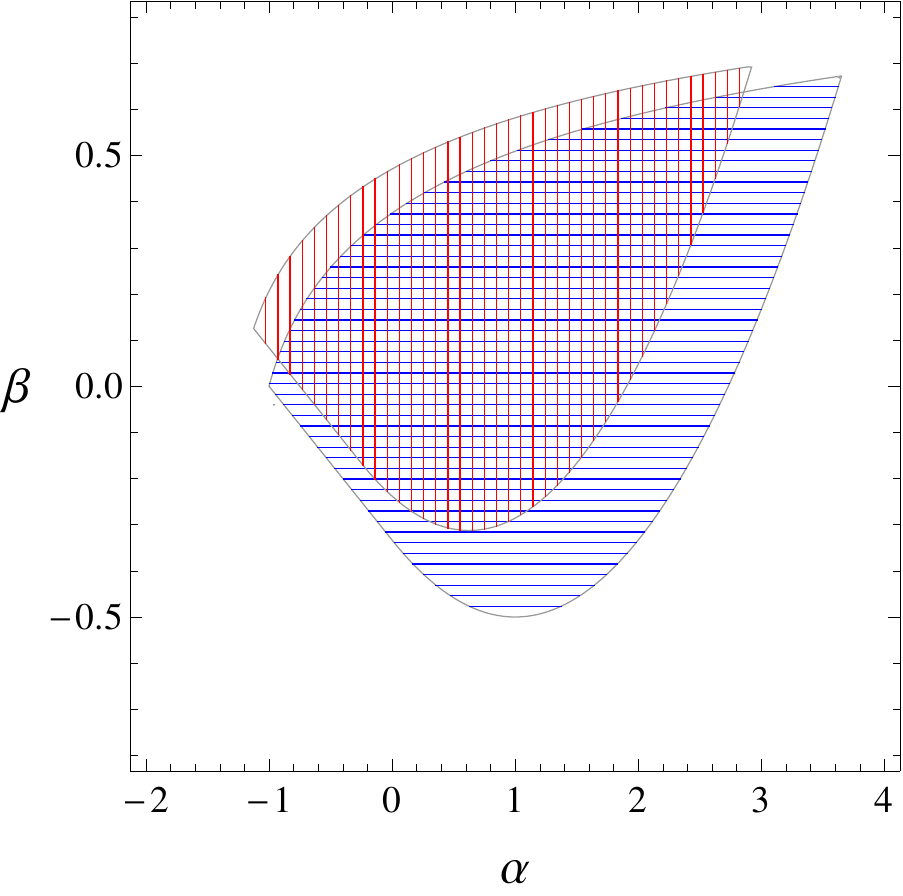}
\caption{The graph in the left panel is the plot of $\alpha$ and $\beta$ which satisfy the stability conditions for the grand canonical ensemble with given values of the parameters; $m_g = 1$, $c=1$, $\mu=0.5$, and $M= 1$. The horizontally-blue-striped region corresponds to the values of $\alpha$ and $\beta$ by which both conditions for local stability in Eq. (\ref{bhcgheatcapcon}) and global stability according to the Eq. (\ref{bhcgHPcon}) are satisfied while the vertically-red-striped region satisfies the conditions $T_{g(\text{min})} > 0$ and $r_+ > 0$. For the right panel, the vertically-red-striped region corresponds to the overlapping region from the left panel,  in which both of the stability conditions and $T_{g(\text{min})} > 0$, $r_+ > 0$ are satisfied,
  compared with that obtained from the non-charged case, which is the blue-shaded horizontally-striped region. }\label{galphabetaplot}
\end{figure}
\begin{figure}[h!]
\includegraphics[scale=0.58]{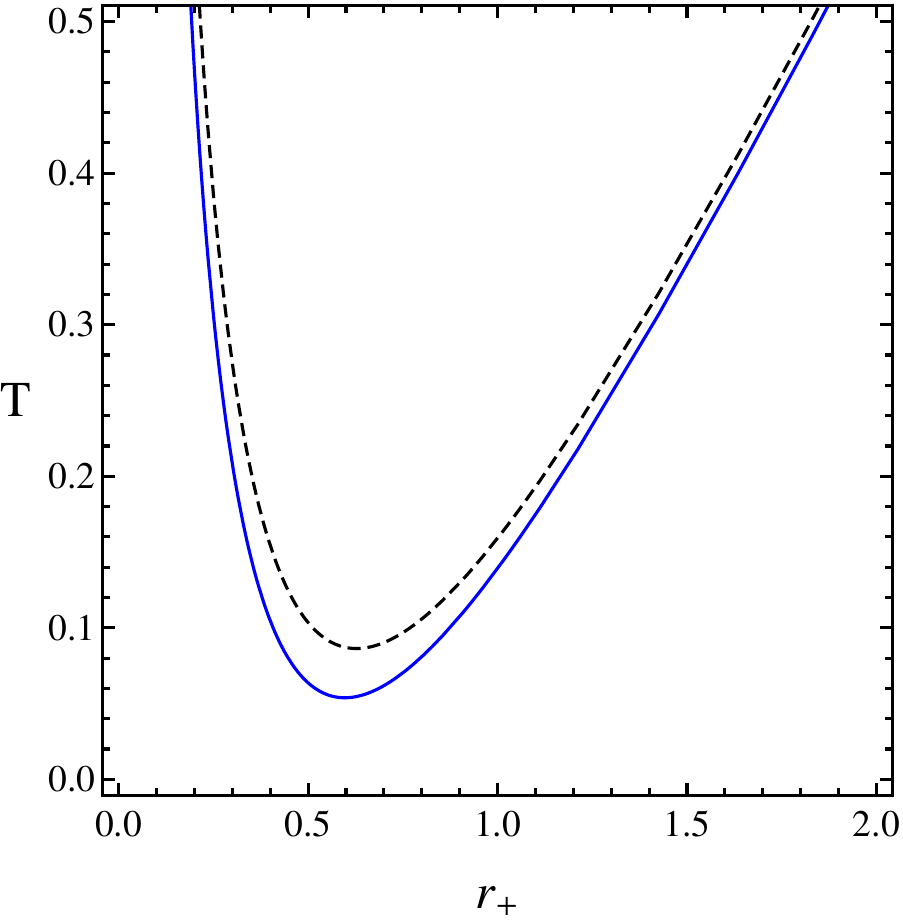}\qquad\quad
\includegraphics[scale=0.58]{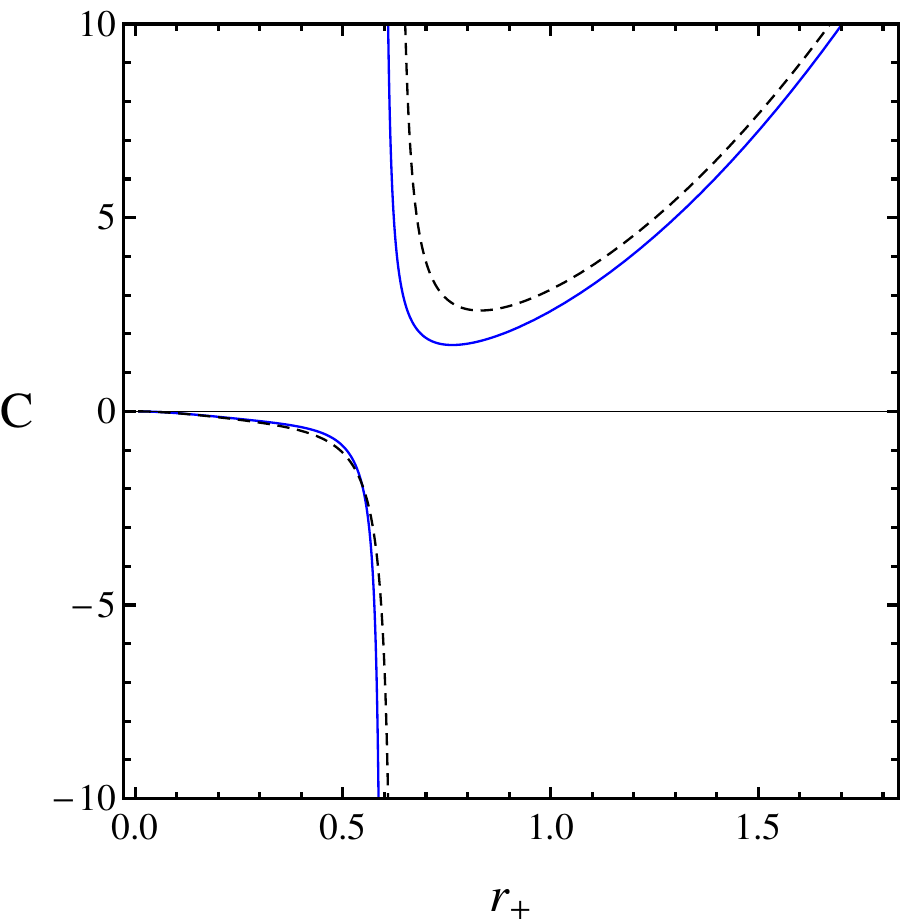}
\caption{These are the plots of the temperature and the heat capacity of the charged black hole respectively where $m_g = 1$, $\alpha =1$, $\beta=0.2$, $c=1$, and $\mu=0.5$. Both the temperature and the heat capacity profiles of the charged black hole are given by the blue thick curves compared with the black dashed curves corresponding to those of the neutral black hole. The divergence in the heat capacity plot corresponds to the minimal temperature of the black hole.}\label{bhcgplotTC}
\end{figure}

For this grand canonical ensemble aspect, both of the stabilities depend  not only on the parameters $\Lambda$ and $\zeta$ like the former analysis but also $\mu$ where the former two are again determined by $\alpha$ and $\beta$. In the left panel of Figure \ref{galphabetaplot}., it shows the validity of those parameters where both the stabilities are assumed. From this figure, one may see that it is not difficult to find the allowed region with parameters $\alpha, \beta \sim O(1)$ which suggests that the theory naturally provides the Hawking-Page phase transition. In the right panel of Figure \ref{galphabetaplot}, a comparison of the allowed regions between charged and non-charged ($\mu=0$) cases is illustrated. From this plot, it is found that the existence of the charge makes the allowed region smaller. Furthermore, we can see the correspondence between the minimal temperature and the divergence of the heat capacity graphically in Figure \ref{bhcgplotTC}.

\subsection{Canonical Ensemble}

\begin{figure}[h!]
\includegraphics[scale=0.6]{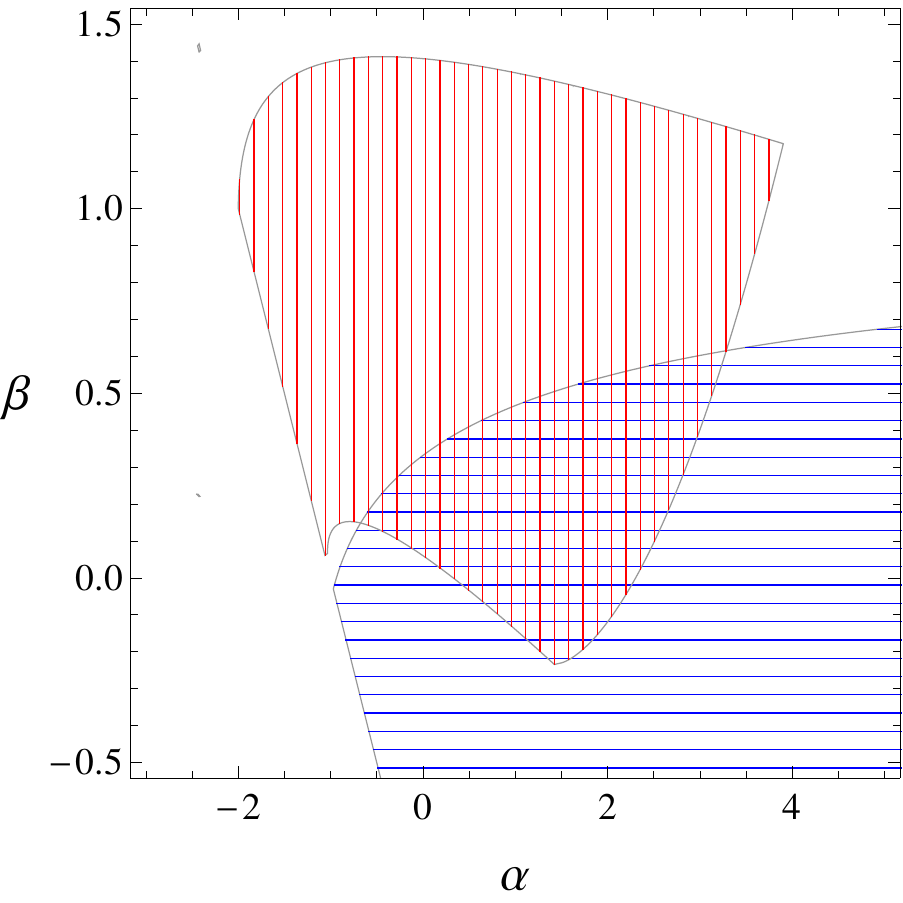}\qquad\quad
\includegraphics[scale=0.6]{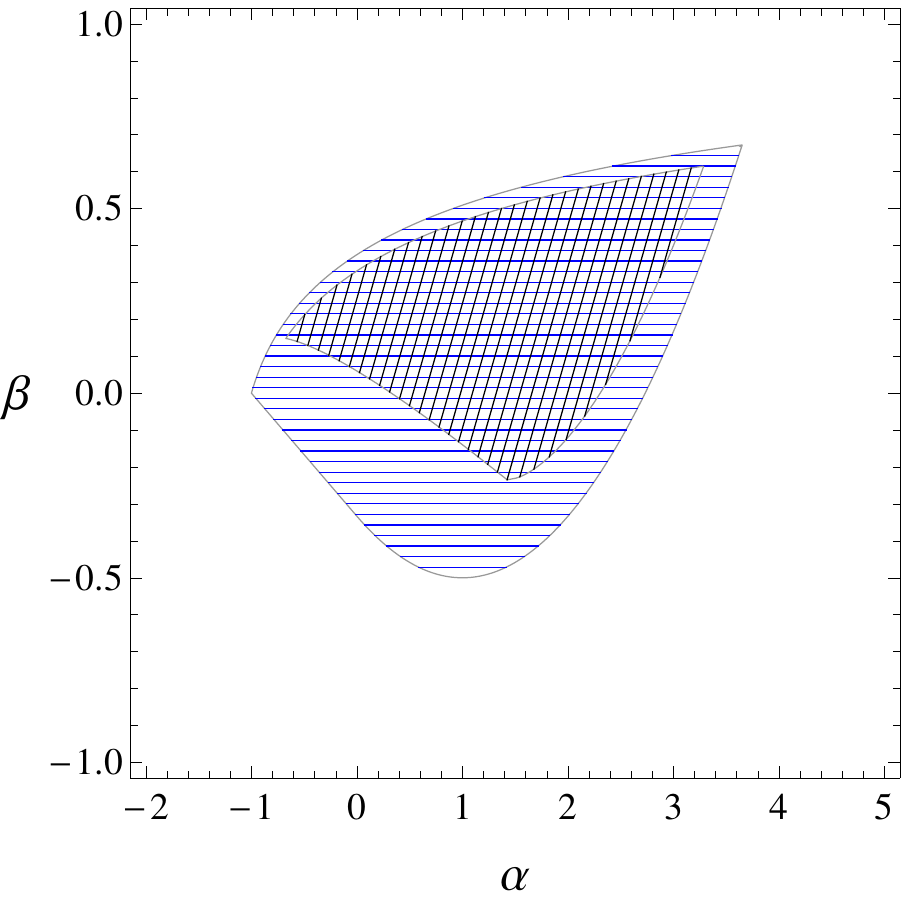}
\caption{The left panel is the region plot in ($\alpha,\beta$) space satisfying both of the stability conditions with given values of the parameters; $m_g = 1$, $c=1$, $Q=0.2$, and $M=1$. The horizontally-blue-striped region corresponds to the values of $\alpha$ and $\beta$ by which both conditions for local stability in Eq. (\ref{bhccheatcapcon}) and global stability according to the Eq. (\ref{bhccHPcon}) are satisfied while the vertically-red-striped region satisfies both of the conditions $T_{c(\text{min})} > 0$ and $r_+ > 0$. For the right panel, the skewed-black-striped illustrates the overlapping region in the left panel, satisfying both local and global stability conditions, $T_{c(\text{min})} > 0$ and $r_+ > 0$ , compared with the one obtained in the non-charged case which is illustrated in the horizontally-blue-striped region.}\label{calphabetaplot}
\end{figure}

\begin{figure}[h!]
\includegraphics[scale=0.6]{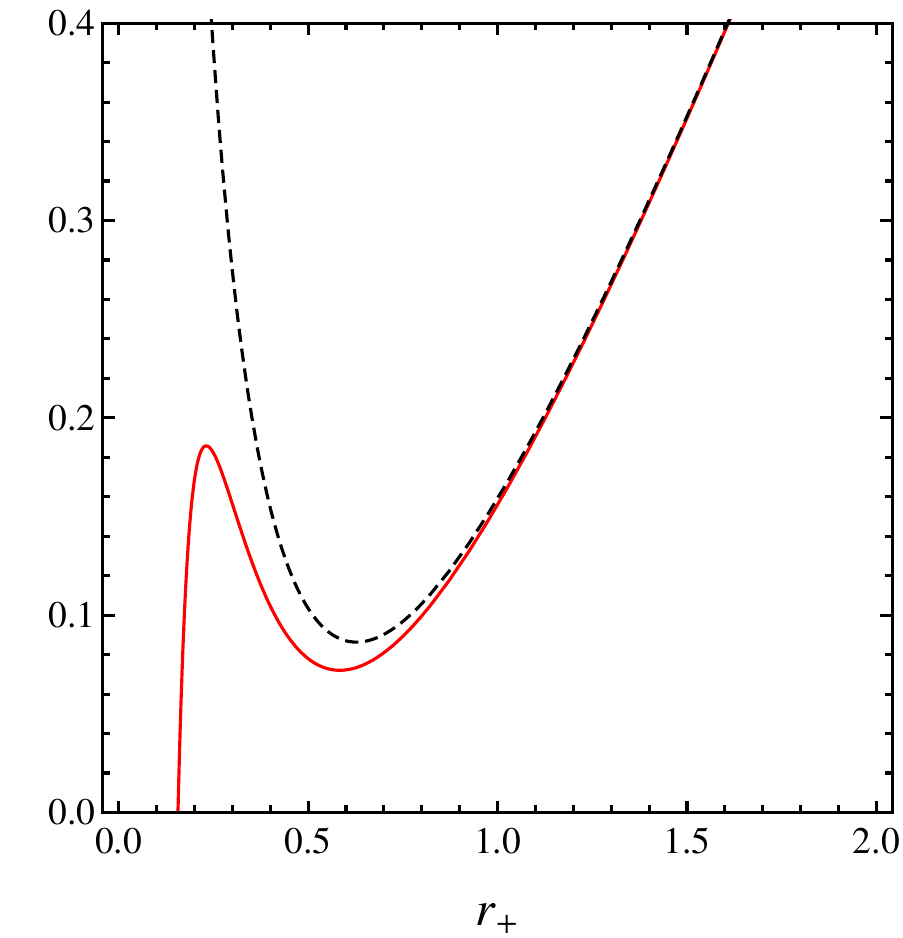}\qquad\quad
\includegraphics[scale=0.6]{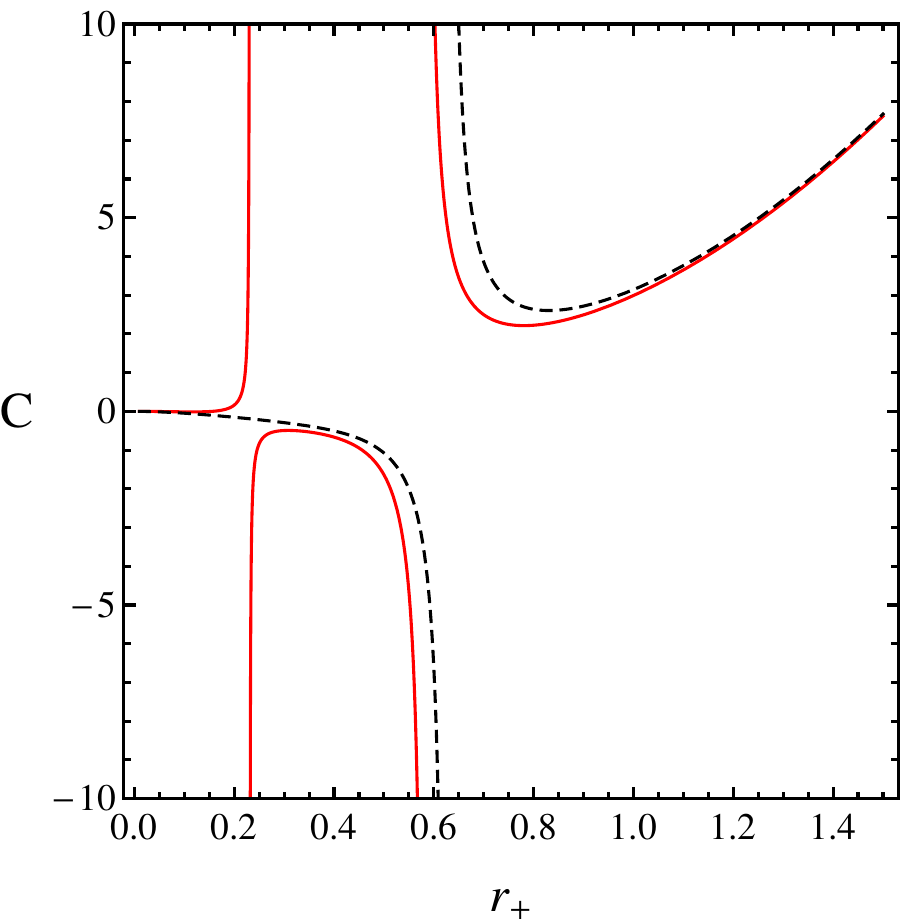}
\caption{These are the plots of the temperature and the heat capacity of the charged black hole respectively in the canonical ensemble view where $m_g = 1$, $\alpha =1$, $\beta=0.2$, $c=1$, and $Q=0.2$. Both the temperature and the heat capacity profiles of the charged black hole are given by the red thick curves compared with the black dashed curves corresponding to those of the neutral black hole. The divergence at large $r_+$ in the heat capacity corresponds to the minimal temperature of the black hole. Moreover, the heat capacity diverges at two values of $r_+$.}\label{bhccplotTC}
\end{figure}

On the other hand, if charge transfer is prohibited, one can consider the black hole as a closed system, or a canonical ensemble with fixed non-zero charge $Q$. The mass and temperature are given readily by  Eq. (\ref{bhcmass}) and Eq. (\ref{bhctemp}) respectively. Furthermore, the corresponding entropy still  obeys the area law,
\begin{eqnarray}
S=\pi r^2_+.
\end{eqnarray}
Since the charge is fixed, the appropriate free energy in consideration is the Helmholtz free energy, which is
\begin{eqnarray}
F&=&M-TS, \nonumber
\\
&=&\frac{1}{4}r_+ \left(1-\frac{\Lambda}{3}r^2_+ +\zeta +3\frac{Q^2}{r^2_+}\right).\label{Helmholtz-FE}
\end{eqnarray}
Similarly, the condition for globally thermodynamical stability is
\begin{eqnarray}
\Lambda r^2_+ \geq 3\left(1 +\zeta +3\frac{Q^2}{r^2_+}\right). \label{bhccHPcon}
\end{eqnarray}
Furthermore, to examine the locally thermodynamical stability, the corresponding heat capacity is computed,
\begin{eqnarray}
C_c=\frac{2\pi r^2_+\left(1+\Lambda r^2_+ +2\gamma r_+ +\zeta -\frac{Q^2}{r^2_+}\right)}{\Lambda r^2_+ -\left(1+\zeta\right)+3\frac{Q^2}{r^2_+}} = \frac{8\pi^2 r^3_+T}{\Lambda r^2_+ -\left(1+\zeta\right)+3\frac{Q^2}{r^2_+}}.
\end{eqnarray}
The condition for a locally stable black hole to exist is
\begin{eqnarray}
\Lambda r^2_+ > \left(1 +\zeta -3\frac{Q^2}{r^2_+}\right). \label{bhccheatcapcon}
\end{eqnarray}
Here, the conditions for both stabilities depend on $\Lambda$ and $\zeta$ (or similarly, $\alpha$ and $\beta$), as well as on the charge $Q$ rather than the potential $\mu$ as in the grand canonical ensemble case, as shown in left panel of Figure \ref{calphabetaplot}. Again, from the allowed region in this figure, the theory can provide the Hawking-Page phase transition naturally. From the right panel of this figure, it is also found that the existence of the charge makes the allowed region smaller, similar to the grand canonical case. This is implied by Eq. (\ref{bhctemp}) where the region for which $T > 0$ is reduced as the presence of charge. Similarly, the correspondence between the minimal temperature and the divergence of the heat capacity can be seen in Figure \ref{bhccplotTC}.

\begin{figure}
\begin{center}
\includegraphics[scale=0.8]{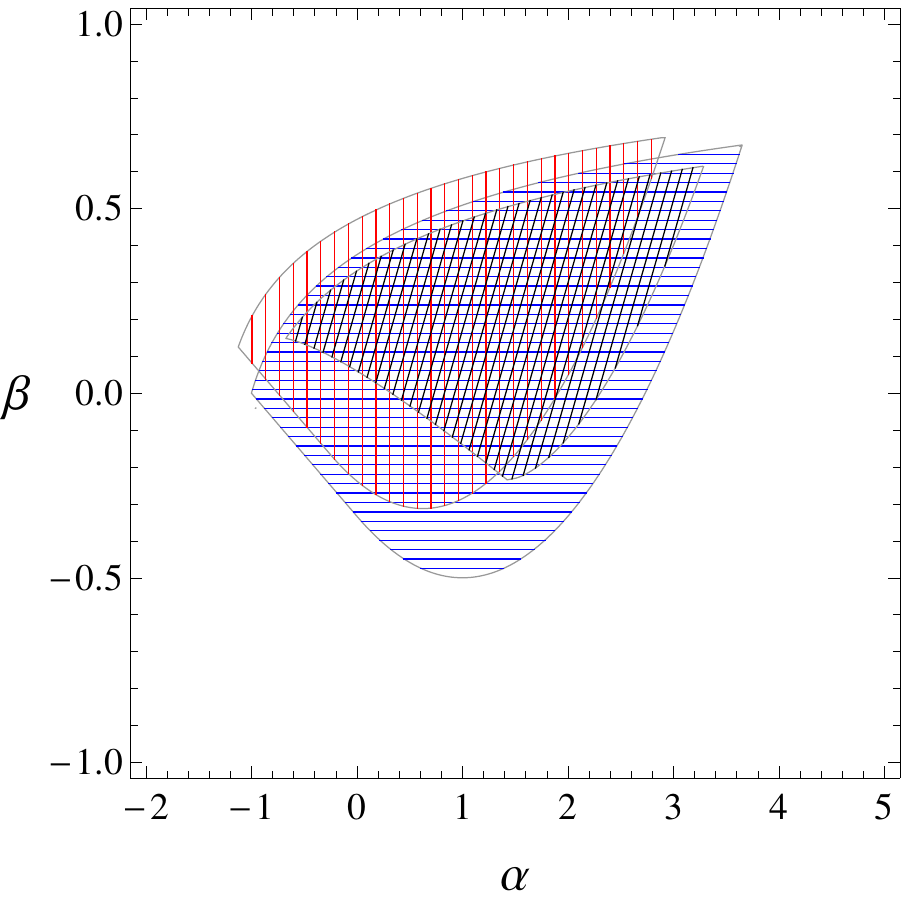}
\end{center}
\caption{This plot shows the valid regions of the parameters $\alpha$ and $\beta$ satisfying both global and local thermodynamical stability as well as the positive temperature condition and $r_+ > 0$  where $m_g=c = 1$, and $M= 1$.  The vertically-red-striped region corresponds to the grand canonical aspect of the black hole with $\mu = 0.5$, the horizontally-blue-striped region corresponds to the neutral black hole, and skewed-black-striped region corresponds to the canonical ensemble one with $Q=0.2$.} \label{HPcompare}
\end{figure}

The effect of charge on the stability regions of both the grand canonical ensemble and the canonical ensemble compared with non-charged case are shown in Figure \ref{HPcompare}. From this figure, one can see the that the effect of the chemical potential $\mu$ in the grand canonical ensemble and the charge $Q$ in the canonical ensemble decreases size of the allowed region of the parameters. These can be seen from  Eq. (\ref{bhctemp}) and Eq. (\ref{bhcgtemp}) since the contribution from the charge and chemical potential makes the positive temperature region smaller. In the canonical ensemble, we also found that the parameter region that satisfies conditions for the existence of four horizons is not compatible with the stability region, similar to the non-charged case. The effect of the parameter $c$ on the region of stability is also similar to the non-charged case. The stability region will increase where the horizon size increases or the parameter $c$ decreases.

\section{Concluding remarks}\label{summary}
The dRGT massive gravity is a natural extension of Einstein's theory of general relativity, providing mass to the graviton, and it is of a great arena for theoretical physics research. The dRGT massive gravity describes nonlinear interaction terms as a correction of the Einstein-Hilbert action and hence admits general relativity as a particular case.  It is believed that dRGT massive gravity may provide a possible explanation for the accelerated expansion of the Universe that does not require any dark energy or cosmological constant. Hence, dRGT massive gravity has received significant attention \cite{deRham:2014zqa} including searches for black holes \cite{msv13}.
In this paper, we have obtained a class of black hole solutions in dRGT massive gravity, and studied the
thermodynamics and phase structure of the black hole solutions.  In the dRGT massive gravity outlined in this paper, there are three terms in the effective potential associated with the graviton mass. Furthermore, we note that due to the inclusion of the massive gravity term in the action, the Schwarzschild solution in the general relativity is modified. Interestingly, it turns out that solutions to the Einstein field equations, such as the monopole-de Sitter-Schwarzschild become solutions in dRGT massive gravity for suitable choices of the parameters of the theory, where the coefficients for the third and fourth terms in the potential and the graviton mass in massive gravity naturally generate the cosmological constant and the global monopole term. The corresponding thermodynamical quantities are also changed.  However, the black hole entropy is not affected significantly by the existence of the graviton mass, and still obeys the standard area law  as in general relativity.  Moreover, we consider the charged black hole solution in both the grand canonical and canonical ensembles to analyze the thermodynamics and phase transition.   We have demonstrated through the calculation of the heat capacity and the free energy that there is a critical point where the heat capacity diverges and a phase transition is possible without requiring fine-tuning of the parameters as shown in Figure \ref{alphabetaplot}. Even though it is possible to obtain three horizons in some region of parameter space, the region is still not compatible with the stability region. This implies that the phase transition will not occur when the black hole has three horizons. The presence of the charge will not change this argument, the phase transition will not occur when the black hole has four horizons. The presence of the charge affects the appearance of the Hawking-Page phase transition such that the allowed parameter region decreases as shown in Figure \ref{HPcompare}. We also found that the phase transition tends to occur in the large horizon size in both the charged and non-charged cases.

The black  hole solutions obtained are immensely simplified due to the choice of the fiducial metric  as
$f_{\mu\nu}= \text{diag}(0,0,c^2  ,c^2 \sin^2\theta)$ and the choice of the St\"uckelberg scalars. It will be interesting to apply the technique discussed here in other massive gravities to get black holes.  It will also  be interesting to consider the motion of particles in the background of the  dRGT massive black holes considered and to see how the graviton mass effects the equations of motion.  These and related areas are for future investigation.

\begin{acknowledgements}
P.W. and L.T. are supported by the Thailand Toray Science Foundation (TTSF) science and technology research grant. P.W is also supported by the Naresuan University Research Fund through grant No. R2557C083.  S.G.G., would like to thank SERB-DST, government of India for Research Project Grant No.  SB/S2/HEP-008/2014. Moreover, this project is partially supported by the ICTP through grant  No. OEA-NET-76. Furthermore, we would like to thank the Institute for Fundamental Study (IF), Naresuan University for hospitalities during the process of this work. Last but not least, we would like to thank Matthew James Lake for reading through the manuscript and correcting some grammatical errors.
\end{acknowledgements}

\begin{appendix}
\section{Alternative form of dRGT action}\label{A.A}
There is one of the alternative form of 4-dimensional dRGT massive gravity which is worthy of discussion \cite{deRham:2014zqa}. To consider the alternative form, it is useful to write the ghost-free massive gravity action in a more general form. In an arbitrary number of  dimensions, $n$, the ghost-free interaction can be constructed as
\begin{eqnarray}
{\cal U}(\mathcal{K}) = \sum^n_{i=0}\alpha_i{\cal U}_i(\mathcal{K}),
\end{eqnarray}
where  the $i^\text{th}$ term corresponds to the  anti-symmetric contraction of the $i^\text{th}$ order of $\mathcal{K}^\mu_\nu$ as follow,
\begin{subequations}
\begin{eqnarray}
\mathcal{U}_0(\mathcal{K}) &\equiv& 1,
\\
\mathcal{U}_1(\mathcal{K}) &\equiv&\mathcal{K}^{\mu_1}_{\mu_1},
\\
\mathcal{U}_2(\mathcal{K}) &\equiv&\mathcal{K}^{\mu_1}_{[\mu_1} \mathcal{K}^{\mu_2}_{\mu_2]},
\\
\mathcal{U}_3(\mathcal{K}) &\equiv&\mathcal{K}^{\mu_1}_{[\mu_1} \mathcal{K}^{\mu_2}_{\mu_2} \mathcal{K}^{\mu_3}_{\mu_3]},
\\
\mathcal{U}_4(\mathcal{K}) &\equiv&\mathcal{K}^{\mu_1}_{[\mu_1} \mathcal{K}^{\mu_2}_{\mu_2} \mathcal{K}^{\mu_3}_{\mu_3} \mathcal{K}^{\mu_4}_{\mu_4]},
\\
&\vdots&
\\
\mathcal{U}_n(\mathcal{K}) &\equiv&\mathcal{K}^{\mu_1}_{[\mu_1} \mathcal{K}^{\mu_2}_{\mu_2} \mathcal{K}^{\mu_3}_{\mu_3} \ldots \mathcal{K}^{\mu_n}_{\mu_n]},
\end{eqnarray}\label{ninteract}
\end{subequations}
where the building block tensor is
\begin{eqnarray}
\mathcal{K}^\mu_\nu=\delta^\mu_\nu-\sqrt{g^{\mu\sigma}f_{ab} \partial_\sigma \phi^a \partial_\nu \phi^b}.
\end{eqnarray}
Note that here the square brackets acting on indices denote anti-symmetrization,
\begin{eqnarray}
[~,~] \equiv \frac{1}{n!}\{(\text{even permutation of n indices})-(\text{odd permutation of n indices})\}.
\end{eqnarray}
Generally speaking, this form of interaction can be expressed in an alternative form with a different definition of the building block tensor, namely
\begin{eqnarray}
{\cal U}(\mathbb{X}) = \sum^n_{i=0}c_i{\cal U}_i(\mathbb{X}).
\end{eqnarray}
where
\begin{eqnarray}
\mathbb{X}=\sqrt{g^{\mu\sigma}f_{ab} \partial_\sigma \phi^a \partial_\nu \phi^b},
\end{eqnarray}

For the 4-dimensional ghost-free massive gravity, the anti-symmetric contractions of the terms that are higher than  $4^\text{th}$ order vanish which leaves the non-zero interaction term in Eq. (\ref{ninteract}) as follows
\begin{subequations}
\begin{eqnarray}
\mathcal{U}_0(\mathcal{K}) &\equiv& 1,\\
\mathcal{U}_1(\mathcal{K}) &\equiv&[\cal{K}],\\
 {\cal U}_2(\mathcal{K})&\equiv&[{\cal K}]^2-[{\cal K}^2] ,\\
 {\cal U}_3(\mathcal{K})&\equiv&[{\cal K}]^3-3[{\cal K}][{\cal K}^2]+2[{\cal K}^3] ,\\
 {\cal U}_4(\mathcal{K})&\equiv&[{\cal K}]^4-6[{\cal K}]^2[{\cal K}^2]+8[{\cal K}][{\cal
K}^3]+3[{\cal K}^2]^2-6[{\cal K}^4],
\end{eqnarray}
\end{subequations}
To transform these terms to the different convention, introduced above, one can transform the coefficients $\alpha_i$'s to $c_i$'s via the transformation matrix,
\begin{eqnarray}
\left(\begin{array}{c}
c_0\\c_1\\c_2\\c_3\\c_4
\end{array}\right) =
\left(
\begin{array}{ccccc}
 1 & 4 & 12 & 24 & 24 \\
 0 & -1 & -6 & -18 & -24 \\
 0 & 0 & 1 & 6 & 12 \\
 0 & 0 & 0 & -\frac{2}{3} & -\frac{2}{3} \\
 0 & 0 & 0 & 0 & \frac{1}{24}
\end{array}
\right)
\left(\begin{array}{c}
\alpha_0\\\alpha_1\\\alpha_2\\\alpha_3\\\alpha_4
\end{array}\right),
\end{eqnarray}
or conversely via the inverse transformation matrix,
\begin{eqnarray}
\left(\begin{array}{c}
\alpha_0\\\alpha_1\\\alpha_2\\\alpha_3\\\alpha_4
\end{array}\right) =
\left(
\begin{array}{ccccc}
 1 & 4 & 12 & 36 & -1152 \\
 0 & -1 & -6 & -27 & 720 \\
 0 & 0 & 1 & 9 & -144 \\
 0 & 0 & 0 & -\frac{3}{2} & -24 \\
 0 & 0 & 0 & 0 & 24
\end{array}
\right)
\left(\begin{array}{c}
c_0\\c_1\\c_2\\c_3\\c_4
\end{array}\right).
\end{eqnarray}
In the case of 4-dimensional dRGT massive gravity, the zeroth order term; $U_0$, corresponds to the cosmological constant, then one can set $\alpha_0 = 0$ for simplicity. Moreover, the tadpole term $U_1$ must vanishes to recover the Fierz-Pauli massive gravity in a linearized level \cite{Fierz:1939ix}, then $\alpha_1=0$. Thus, the 4-dimensional dRGT massive gravity can be expressed in an alternative form as
\begin{subequations}
\begin{eqnarray}
c_0&=&12\left(1+2\alpha_3+2\alpha_4\right),\\
c_1&=&-6\left(1+3\alpha_3+4\alpha_4\right),\\
c_2&=&1+6\alpha_3+12\alpha_4~,\\
c_3&=&-\frac{2}{3}\left(\alpha_3+\alpha_4\right),\\
c_4&=&\frac{\alpha_4}{24}.
\end{eqnarray}\label{alter1}
\end{subequations}
In addition, with the redefinition of parameters in Eq. (\ref{alphabeta}), one can rewrite Eq. (\ref{alter1}) as
\begin{subequations}
\begin{eqnarray}
c_0&=&6\left(1+\alpha+\beta\right),\\
c_1&=&-2\left(1+2\alpha+3\beta\right),\\
c_2&=&\alpha+3\beta~,\\
c_3&=&\frac{1}{6}\left(1-\alpha-\beta\right),\\
c_4&=&\frac{1}{288}\left(1-\alpha+3\beta\right).
\end{eqnarray}\label{alter2}
\end{subequations}
Moreover, the black hole solution in Eq. (\ref{solutionf}) can be rewritten in this convention as
\begin{subequations}
\begin{eqnarray}
f(r) =1 - \frac{2 M}{r}+ \frac{\Lambda}{3}r^2+\gamma r+\zeta , \nonumber
\end{eqnarray}
where, alternatively,
\begin{eqnarray}
\Lambda&=& \frac{m^2_g}{2} c_0 ,\\
\gamma&=&\frac{c m^2_g}{2} c_1 ,\\
\zeta&=&c^2 m^2_g c_2 .
\end{eqnarray}
\end{subequations}
\end{appendix}

\end{document}